\documentclass[reprint,amsmath,amssymb, aps,pra]{revtex4-2}

\usepackage{graphicx}
\usepackage{dcolumn}
\usepackage{bm}
\usepackage{hyperref}
\usepackage{chngcntr}
\usepackage[colorinlistoftodos]{todonotes}
\usepackage{adjustbox}
\usepackage{amsmath}
\usepackage{siunitx}
\presetkeys{todonotes}{inline,backgroundcolor=yellow}{}
\renewcommand{\thefigure}{\Roman{figure}}

\begin{document}

\title{A Comparative Analysis of Non-relativistic and Relativistic Calculations of Electric Dipole Moments and Polarizabilities of Heteronuclear Alkali Dimers}

\author{R. Mitra}
\affiliation{Physical Research Laboratory, Atomic, Molecular and Optical Physics Division, Navrangpura, Ahmedabad-380009, India and \\
Indian Institute of Technology Gandhinagar, Palaj, Gandhinagar-382355, Gujarat, India}
\author{V. S. Prasannaa}
\email{srinivasaprasannaa@gmail.com}
\affiliation{Physical Research Laboratory, Atomic, Molecular and Optical Physics Division, Navrangpura, Ahmedabad-380009, India}
\author{B. K. Sahoo}
\affiliation{Physical Research Laboratory, Atomic, Molecular and Optical Physics Division, Navrangpura, Ahmedabad-380009, India}

\begin{abstract}
We analyze the molecular electric dipole moments (PDMs) and static electric dipole polarizabilities of heteronuclear alkali dimers in their ground states by employing coupled-cluster theory, both in the non-relativistic and four-component relativistic frameworks. The roles of electron correlations as well as relativistic effects are demonstrated by studying them at different levels of theory, followed by a comprehensive treatment of error estimates. We compare our obtained values with the previous non-relativistic calculations, some of which include lower-order relativistic corrections, as well as with the experimental values, wherever available. We find that the PDMs are very sensitive to relativistic effects, as compared to polarizabilities; this aspect can explain the long-standing question on the difference between experimental values and theoretical results for LiNa. We show that consideration of relativistic values of PDMs improves significantly the isotropic Van der Waals $C_6$ coefficients of the investigated alkali dimers over the previously reported non-relativistic calculations. The dependence of dipole polarizabilites on molecular volume is also illustrated. 
\end{abstract}

\maketitle

\section{Introduction} \label{sec1}

In recent years, there has been tremendous interest in the field of ultracold molecules due to their wide array of applications~\cite{Jin,uses} including exciting possibilities such as probing fundamental constants~\cite{VVF}. A notable molecular property that plays a key role in several of these applications is the permanent electric dipole moment (PDM) of a molecule. Molecules with fairly large PDMs give rise to long-range and anisotropic dipole-dipole interactions that can be controlled by external electric fields~\cite{Lem}. A large PDM implies that a sufficiently low external electric field can align species of a molecule for realizing the dipole-dipole interactions~\cite{Molony}. In fact, a knowledge of PDM would help in understanding the dipole interaction strengths for a given density of trapped molecules~\cite{Loh}. The electric dipole-dipole interactions find their applications in the booming field of quantum phase transitions~\cite{phasetransition}. These interactions could couple qubits, which are described as molecular electric dipoles along or against an external electric field, thus opening up avenues for quantum computation with trapped polar molecules~\cite{DeMille,Rabi,SDthesis}. The PDM plays an important role in chaining of polar molecules. It is predicted in Ref.~\cite{chain} that the interaction strengths for this process for molecules in a one-dimensional optical lattice are directly proportional to the square of the PDM. The static dipole polarizability is also an important property in the field of ultracold physics. The restoring force of a trapping laser beam is proportional to the static dipole polarizability of the molecule. Hence, higher the value of dipole polarizability, more the suitability of the molecule for trapping and laser cooling. When molecules are trapped in a far-off resonance optical trap, the static polarizability helps to predict the depth of the trap depending on the intensity of the laser fields~\cite{Loh}. Moreover, polarizability plays a role in femtosecond spectroscopy, specifically in laser-induced impulsive alignment of molecules~\cite{Ren}. Therefore, knowledge of both PDMs and dipole polarizabilities are important for studying ultracold molecules trapped in laser fields \cite{1}. Accurate estimates of both the properties become very relevant for heavier molecules for which not many previous works are available. Of late, heteronuclear alkali dimers have been successfully produced in large numbers in experiments (for example, see Ref.~\cite{Feshbach} and {\it references therein}). Some of the more recent works that realized these molecules either by Feshbach resonance or photoassociation include  Refs.~\cite{Molony,Ye,a,b,c,d,e,f}. The sheer number of experiments make these molecules attractive for several applications, such as quantum information technology, quantum simulations of condensed phase physics, studies of chemical reactions etc. \cite{uses}. For instances, in Ref.~\cite{Buchler}, the authors studied three-body interactions in polar molecules and undertook LiCs for investigation due to its large PDM. In fact, one can view the prospects of orienting and aligning alkali dimers in terms of their PDMs and polarizabilities, respectively~\cite{Deiglmayr}. The  importance of PDMs and polarizabilities, especially in the context of alkali dimers, are further discussed in Ref.~\cite{review}. 
\begin{table*}[t]
\centering
    \caption{The PDM ($\mu$) values (in a.u.) of the investigated alkali dimers. We compare these values from various calculations and available experimental results. Our results from both the non-relativistic and the relativistic methods are given separately. The errors are quoted within the parentheses. }
    \label{tab:table1}
\begin{tabular}{lcccccccccc}
\hline \hline
Method&LiNa&LiK&LiRb&LiCs&NaK&NaRb&NaCs&KRb&KCs&RbCs\\
 \hline \\
 \multicolumn{11}{c}{\textbf{PDM results}}\\
\multicolumn{11}{l}{\textbf{This work}}\\
HF &0.26&1.65&2.09&2.92&1.41&1.88&2.75&0.50&1.46&0.98\\
CCSD &0.25&1.49&1.88&2.70&1.19&1.57&2.39&0.42&1.28&0.87\\
CCSD(T) &0.23&1.39&1.75&2.54&1.09&1.43&2.21&0.36&1.14&0.77\\
\\
DF&0.25&1.62&1.96&2.61&1.39&1.75&2.42&0.37&1.10&0.73\\
RCCSD &0.24&1.45&1.72&2.33&1.16&1.43&2.02&0.28&0.90&0.61\\
RCCSD(T) &0.22&1.36&1.59&2.16&1.07&1.29&1.83&0.24&0.78&0.53\\
RCCSD(T): QZ &0.197&&&&&&&&&\\
RCCSD(T): CBS &0.178&&&&&&&&&\\ \\
\multicolumn{11}{l}{\textbf{Previous calculations}}\\
CCSD(T)~\cite{Urban}&0.17&1.36&1.71&&1.12&1.43&&0.43&&\\
CI~\cite{1999}&0.19&&&&&&&&&\\
CI~\cite{M. Aymar}: Basis A&0.22&1.40&1.64&2.17&1.09&1.30&1.82&0.24&0.75&0.49\\
CI~\cite{M. Aymar}: Basis B&0.22&1.39&1.63&2.17&1.08&1.30&1.83&0.23&0.76&0.50\\
CI~\cite{M. Aymar}: Basis C&&&&2.15&&&1.80&&0.72&0.47\\
CI~\cite{Vanner}&0.23&&&&&&&&\\
CCSD(T)~\cite{Zuchowski}&0.19&1.34&1.57&2.12&1.07&1.30&1.82&0.24&0.78&0.52\\
CCSDT~\cite{Dmitry A. Fedorov}&0.21&1.34&1.60&2.11&1.05&1.29&1.78&0.26&0.75&0.48\\ \\
\multicolumn{11}{l}{\textbf{Experiment}}\\
Ref.~\cite{P. J. Dagdigian}&0.18(1)&&&&&&&&\\
Ref.~\cite{Dag2}&0.1822(7)&1.36(4)&1.57(4)&&1.09(4)&1.22(12)&1.87(8)&&\\
Ref.~\cite{Graff}&0.18&&&&&&&&&\\
Ref.~\cite{F. Engelke}&0.1777(2)&&&&&&&&\\
Ref.~\cite{F. Engelek}&&1.381(2)&&&&&&&\\
Ref.~\cite{Tarnovsky}*&0.18&1.52&1.59&2.48&1.35&1.38&2.31&0.08&1.02&0.94\\
Ref.~\cite{Ye}&&&&&&&&0.2227(8)&&\\
Ref.~\cite{RbCs}&&&&&&&&&&0.51(4)\\
\hline \hline
    \end{tabular}
    \\
    *The values given for a molecule $XY$ that is made of atoms $X$ and $Y$ are actually obtained by employing an empirical rule, which requires a combination of experimental values of polarizabilities of the homonuclear $X_2$ and $Y_2$ molecules, and the values of PDM from the then-recent literature.  
\end{table*}

Only a handful of experimental values for PDMs of the alkali dimers have been reported in literature~\cite{P. J. Dagdigian,Dag2,Graff,F. Engelke,F. Engelek,Tarnovsky,Ye,RbCs}. Experimental data is more scarce for the dipole polarizabilities of these molecules~\cite{Tarnovsky,Graff}. On the other hand, there are numerous  calculations for the PDMs available by employing variants of many-body theories, from as early as 1970s (e.g., see Ref.~\cite{S. Green}) until very recently~\cite{M. Aymar,Deiglmayr,Zuchowski,Dmitry A. Fedorov}. Polarizabilities have not been explored as much theoretically, but a few studies have been carried out on this property ~\cite{Urban,Deiglmayr,Zuchowski}. However, these calculations were performed by using non-relativistic  methods, with some works including lower-order  relativistic corrections \cite{M. Aymar,Deiglmayr,Zuchowski,Dmitry A. Fedorov}. In the heavier alkali dimers, the orbitals get deformed more prominently due to the relativistic effects. Hence, we expect significant deviations from the non-relativistic values for the PDMs and static dipole polarizabilities. Earlier, Lim \textit{et al} \cite{Lim} had investigated static dipole polarizabilities of homonuclear alkali dimers, and found that relativistic effects become important for the heavier dimers. In their calculations, relativistic effects were mainly included through scalar 2-component Douglas-Kroll (DK) Hamiltonian. 

\begin{figure*}[t]
\centering
\begin{tabular}{ccc}
\includegraphics[width=5cm, height=4.5cm]{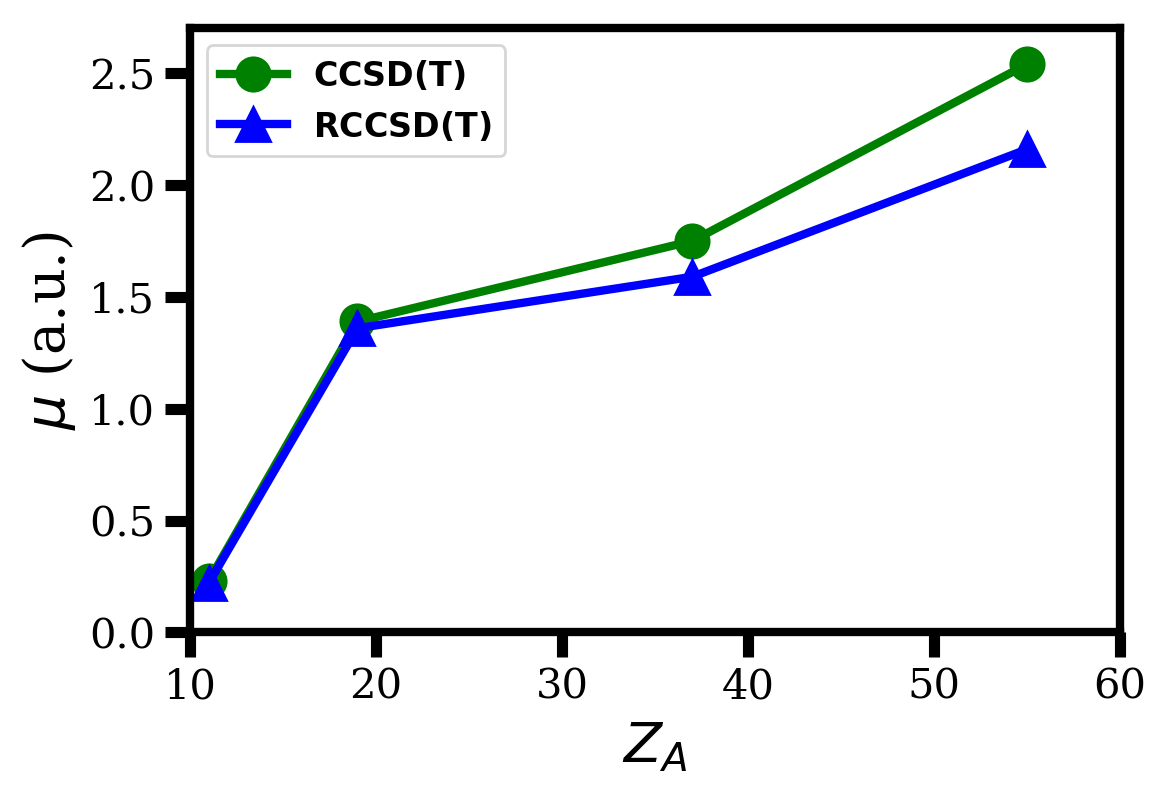} & \includegraphics[width=5cm, height=4.5cm]{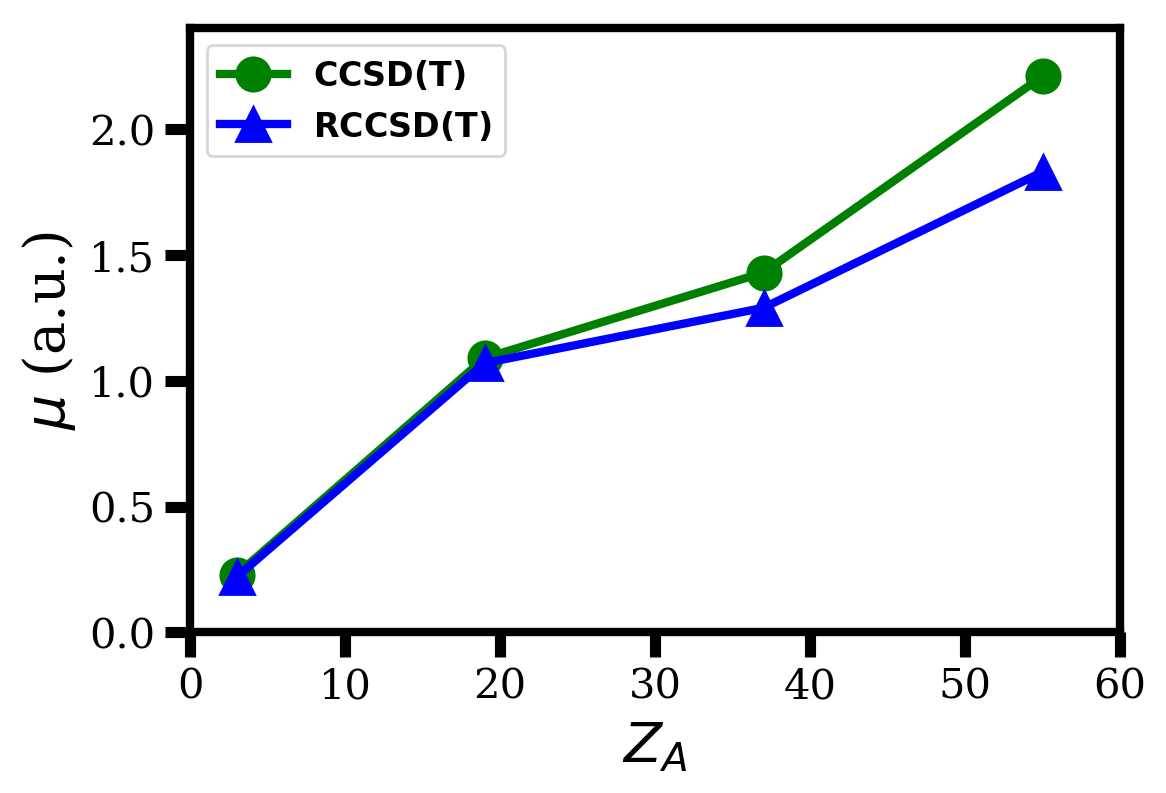} & \includegraphics[width=5cm, height=4.5cm]{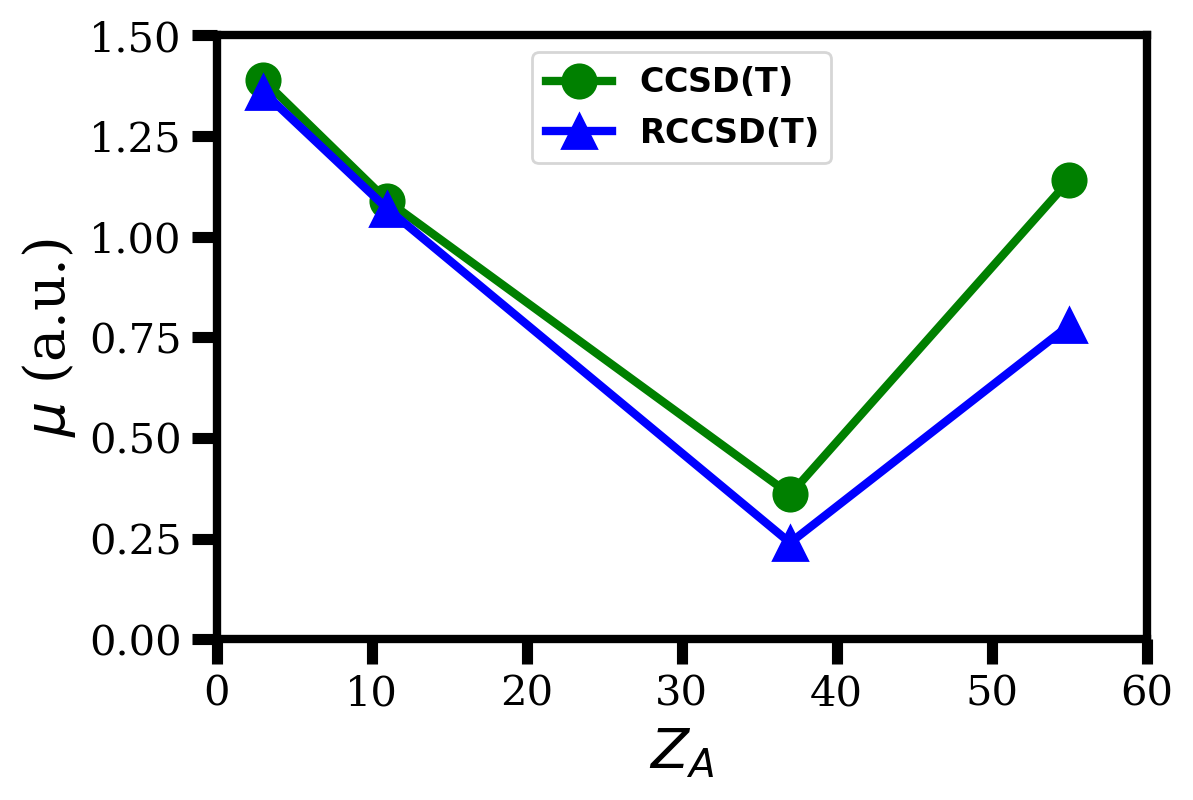}  \\
 (a) & (b) & (c) \\
\includegraphics[width=5cm, height=4.5cm]{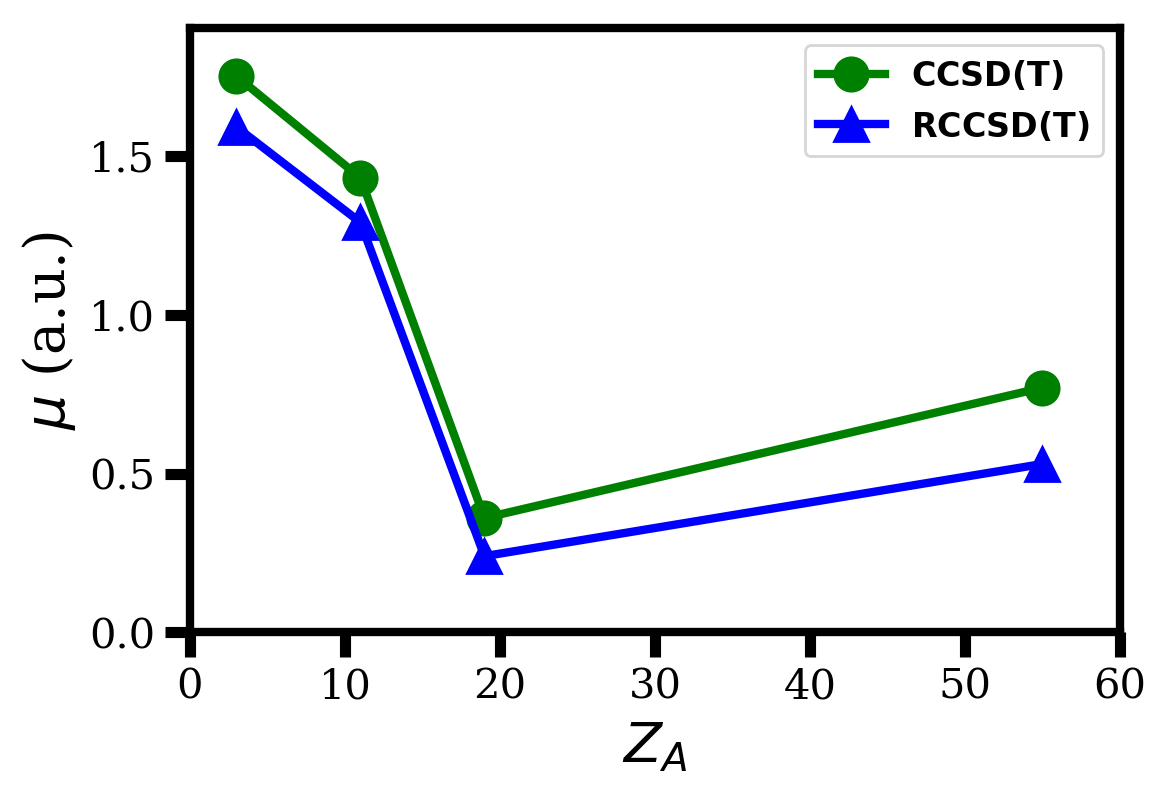} & \includegraphics[width=5cm, height=4.5cm]{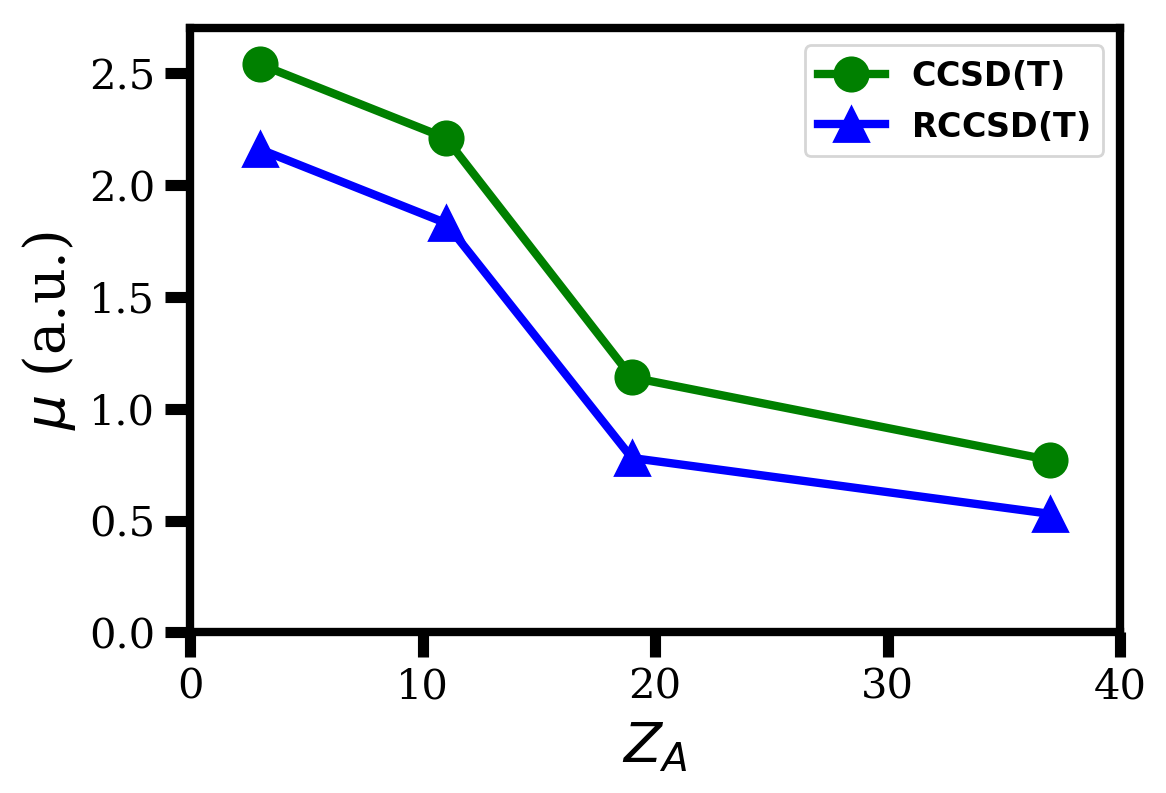} & \\
(d) & (e) & \\
\end{tabular}
\caption{(Colour online) Plots comparing the CCSD(T) and RCCSD(T) values of $\mu$ (in a.u.) for heteronuclear alkali dimers belonging to the (a) Li-, (b) Na-, (c) K-, (d) Rb-, and (e) Cs-families against the atomic number $Z_A$. The trends are different for each family. As expected, relativistic effects are seen to be quite prominent in the heavier molecules.}
\label{fig:figure1}
\end{figure*}

In our work, we investigate the roles of relativistic effects in the PDMs  and dipole polarizabilities of heteronuclear alkali dimers, made of Li, Na, K, Rb and Cs. For this purpose, we perform calculations by employing a non-relativsitic Hamiltonian and 4-component Dirac-Coulomb Hamiltonian in the (relativistic) coupled-cluster ((R)CC) theory. We have adopted the finite field (FF) approach to estimate the first-order and the second-order perturbed energies of the ground states of the above heteronuclear alkali dimers, by varying an electric field. From these energies, we infer the values of the PDMs and dipole polarizabilities. The results are first obtained using the Hartree-Fock (HF) and Dirac-Fock (DF) methods. Electron correlation effects are then systematically included by employing (R)CC theory in the singles and doubles approximation ((R)CCSD method), followed by (R)CC theory in singles, doubles and partial triples approximation ((R)CCSD(T) method). Our results are compared with the previously reported non-relativistic results as well as those obtained from a 2-component scalar relativistic DK Hamiltonian. We also compare our calculated values with the experimental results, wherever available. In doing so, we investigate the large discrepancies seen earlier between the theoretical and experimental results in the PDM of LiNa, and attempt to resolve it. We verify the variation of the components of polarizability with volume using our relativistic results for polarizabilities. We present detailed estimates of possible errors in our calculations. Finally, we discuss the extent to which accurately evaluated PDMs using a relativistic theory could affect the isotropic $C_6$ coefficients of the intermolecular Van der Waals potential. 

\begin{figure}[t]
\centering
\includegraphics[width=8.5cm, height=6.5cm]{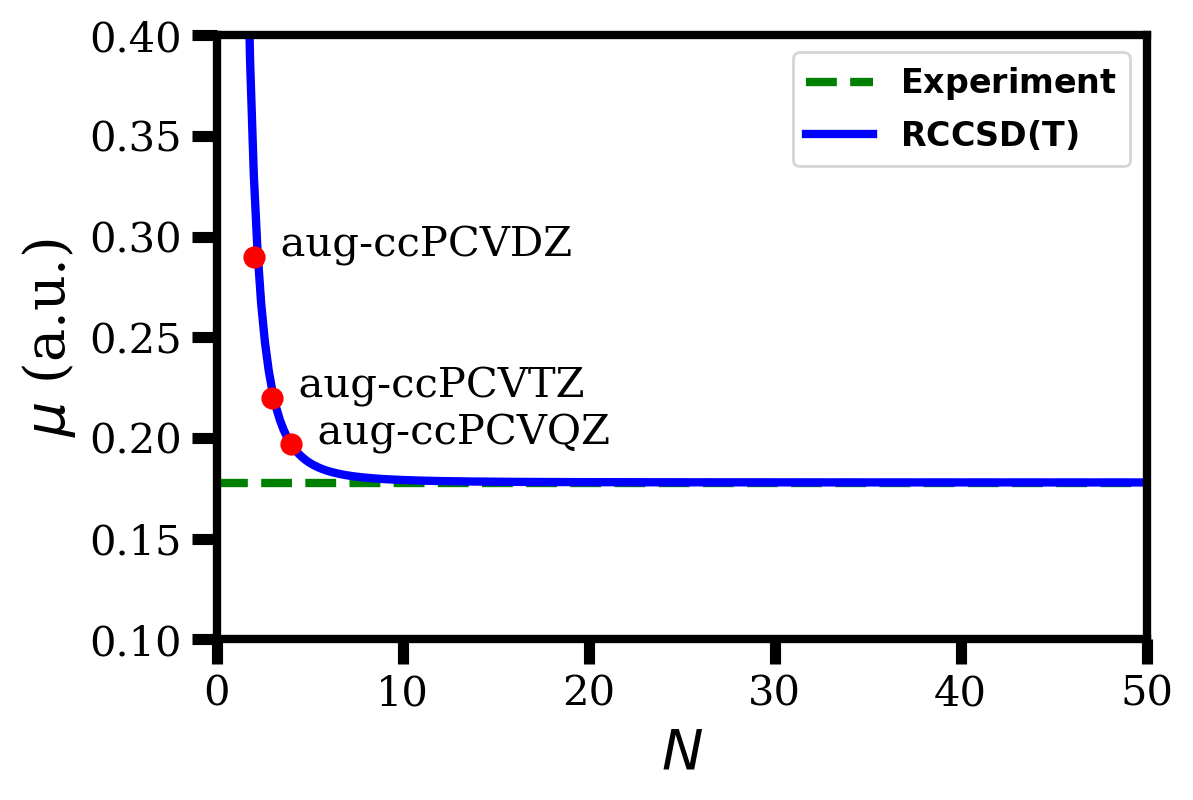}
\caption{(Colour online) Demonstration of the complete basis set extrapolation for the PDM (in a.u.) of LiNa, calculated using the RCCSD(T) method.}
\label{fig:figure2}
\end{figure}

\begin{table*}[t]
    \caption{A comparative analysis of parallel component of the dipole polarizabilities, $\alpha_{\parallel}$ (in a.u.), between the non-relativistic and relativistic calculations. We also present results from the earlier studies. }
    \label{tab:table2}
\begin{tabular*}{\textwidth}{lc@{\extracolsep{\fill}}ccccccccc}
\hline \hline 
Method &LiNa&LiK&LiRb&LiCs&NaK&NaRb&NaCs&KRb&KCs&RbCs\\
\hline
\\
\multicolumn{11}{c}{\textbf{This work}}\\
HF &301.98&425.92&470.95&525.74&502.15&553.93&625.42&816.33&955.87&1084.57\\
CCSD &341.02&470.47&512.88&563.86&522.24&574.12&653.41&792.99&929.26&1023.90\\
CCSD(T) &344.78&482.64&530.04&594.94&531.01&587.90&685.49&794.12&942.39&1025.35\\
\\
DF &301.63&425.52&465.38&529.48&500.20&549.97&621.75&808.95&932.45&1034.89\\
RCCSD &340.55&470.06&507.80&577.23&519.73&567.46&648.24&783.30&900.21&971.17\\
RCCSD(T) &344.29&481.94&523.26&604.40&528.01&578.50&672.12&780.48&902.22&962.85\\
\\
\multicolumn{11}{c}{\textbf{Previous works}}\\
CCSD(T)\cite{Urban}&352.26&484.53&591.83&&537.16&606&&842.19&&\\
CASSCF/NC \cite{2000} &&532.00&&&&&&&&\\
CASSCF/BKPT \cite{2000} &&532.00&&&&&&&&\\
CASPT2/NC \cite{2000} &&512.90&&&&&&&&\\
TDGI \cite{2003} &350.6&&&&&&&&&\\
CCSD(T) \cite{2003} &352.3&&&&&&&&&\\
CI\cite{Deiglmayr}&347.6&489.7&524.3&597.0&529.2&572.0&670.7&748.7&822.3&904.0\\
\hline \hline
\end{tabular*}
\end{table*}

The manuscript is organized as follows: Sec. \ref{sec2} discusses the theory of PDMs and static dipole polarizabilities, and after introducing the (R)CC method, we present the details of obtaining the properties using the FF approach. Sec. \ref{sec3} presents our results and discusses them in detail. We then delve into the trends that we observe for the above properties, with emphasis on the differences between the relativistic and the non-relativistic calculations. We also list and compare our obtained values with the available works in the literature. We then briefly discuss the variation of the components of polarizability with volume. This is followed by a detailed discussion of the possible sources of error, and quote their estimated values. In the last sub-section, we present improved values of $C_6$ coefficients of the alkali dimers. Finally, we conclude in Sec. \ref{sec4}. Unless stated otherwise, atomic units (a.u.) are used throughout the paper. Also, in comparing our results with other works that do not report the results in a.u., we used conversion factors of 1 Debye = 0.3934 a.u. and 1 \AA$^3$ = 6.7483 a.u.$^3$. 

\section{Theory and Methodology} \label{sec2}

In the presence of a weak, static, and homogeneous electric field of strength ${\mathcal E}$, the ground state energy ($E_0$) of a molecule can be expressed as 
\begin{equation}
E_0=E_0^{(0)}+\mathcal{E}E_0 ^{(1)}+\mathcal{E}^2 E_0^{(2)}+ \cdots , \label{100}
\end{equation}
where $E_0^{(0)}$, $E_0^{(1)}$, $E_0^{(2)}$ etc. are the zeroth-order, first-order, second-order etc. contributions to the total energy, respectively. In traditional form, it can be written as
\begin{eqnarray}
E_0 = E_0^{(0)} - \mu_i {\cal E}_i - \frac{1}{2} \alpha_{ij} {\cal E}_i  {\cal E}_j + \cdots ,
\label{eq1}
\end{eqnarray}
where the indices $i$ and $j$ run from 1 to 3, while $\mu_i$ and $\alpha_{ij}$ are the  components of the vector PDM ($\mu$) and rank-two dipole polarizability tensor ($\alpha$), respectively. Now, invoking the Taylor series expansion, it yields
\begin{equation}
E_0=E_0^{(0)}+ \frac{\partial E_0}{\partial\mathcal{E}_i}\Bigg |_{\mathcal{E}_i=0}\mathcal{E}_i + \frac{1}{2!} \frac{\partial^2 E_0}{\partial\mathcal{E}_i\partial\mathcal{E}_j  }\Bigg |_{\genfrac{}{}{0pt}{}{\ \mathcal{E}_i=0,}{\mathcal{E}_j=0}} \mathcal{E}_i\mathcal{E}_j   +\cdots . \label{101}
\end{equation}
Comparing Eqs. (\ref{eq1}) and (\ref{101}), we get
\begin{equation}
\mu_i =-\frac{\partial E_0}{\partial\mathcal{E}_i}\Bigg |_{\mathcal{E}_i=0}     \label{102}
\end{equation}
and
\begin{equation}
\alpha_{ij} = - \frac{\partial^2 E_0}{\partial\mathcal{E}_i\partial\mathcal{E}_j  }\Bigg |_{\genfrac{}{}{0pt}{}{\ \mathcal{E}_i=0,}{\mathcal{E}_j=0}} . \label{103}
\end{equation}
Using these components, the average dipole polarizability ($\bar\alpha$) of a polar molecule is defined as
\begin{eqnarray}
\bar\alpha =\frac{1}{3}(\alpha_{xx}+\alpha_{yy}+\alpha_{zz})=\frac{1}{3}(\alpha_{zz}+2\alpha_{xx}).
\end{eqnarray}
Here, the quantization axis is assumed along the bond length and is in the z- direction. Therefore, it follows that $\alpha_{xx}=\alpha_{yy}$, leading to the last part of the above equation. It is common to denote $\alpha_{zz}$ as $\alpha_{\parallel}$, and $\alpha_{xx}$ and $\alpha_{yy}$ as $\alpha_{\perp}$, for such diatomic systems. Hence, 
\begin{eqnarray}
\bar\alpha =\frac{1}{3}(\alpha_{\parallel}+2\alpha_{\perp}).
\end{eqnarray}

\begin{figure*}[t]
\centering
\begin{tabular}{ccc}
\includegraphics[width=5cm, height=4.5cm]{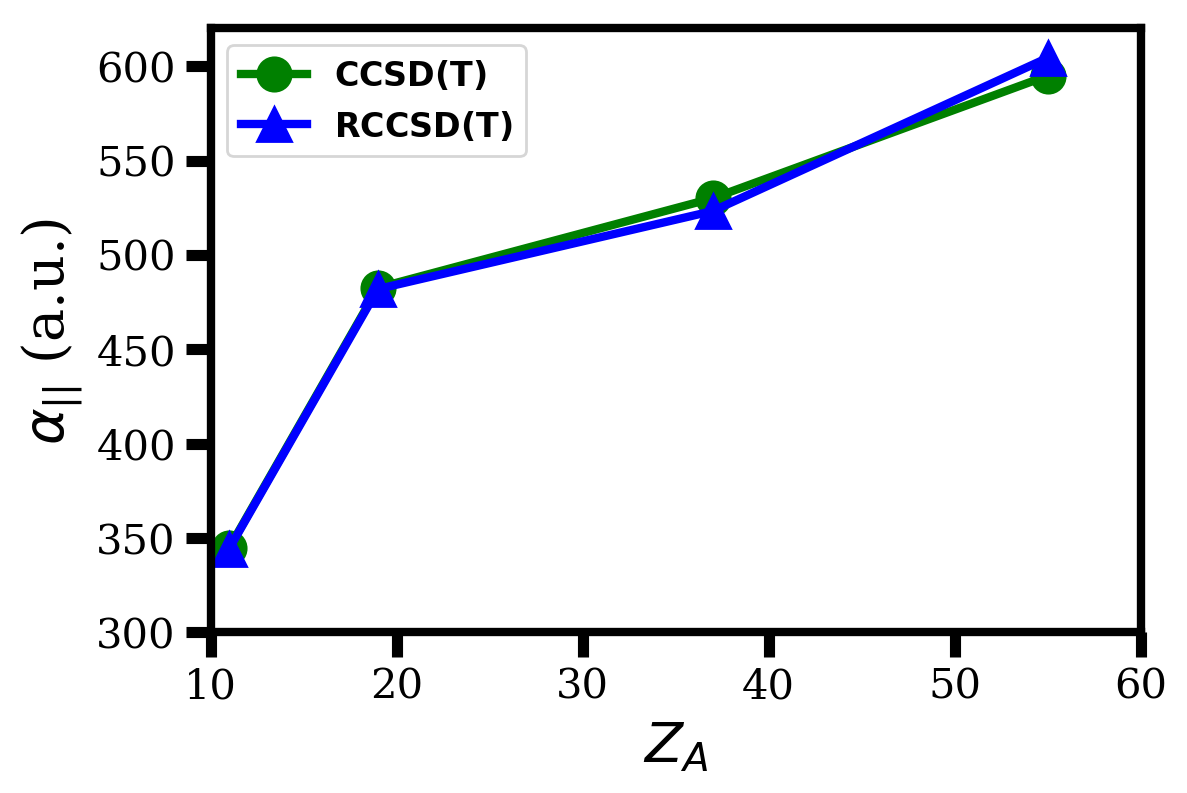} & \includegraphics[width=5cm, height=4.5cm]{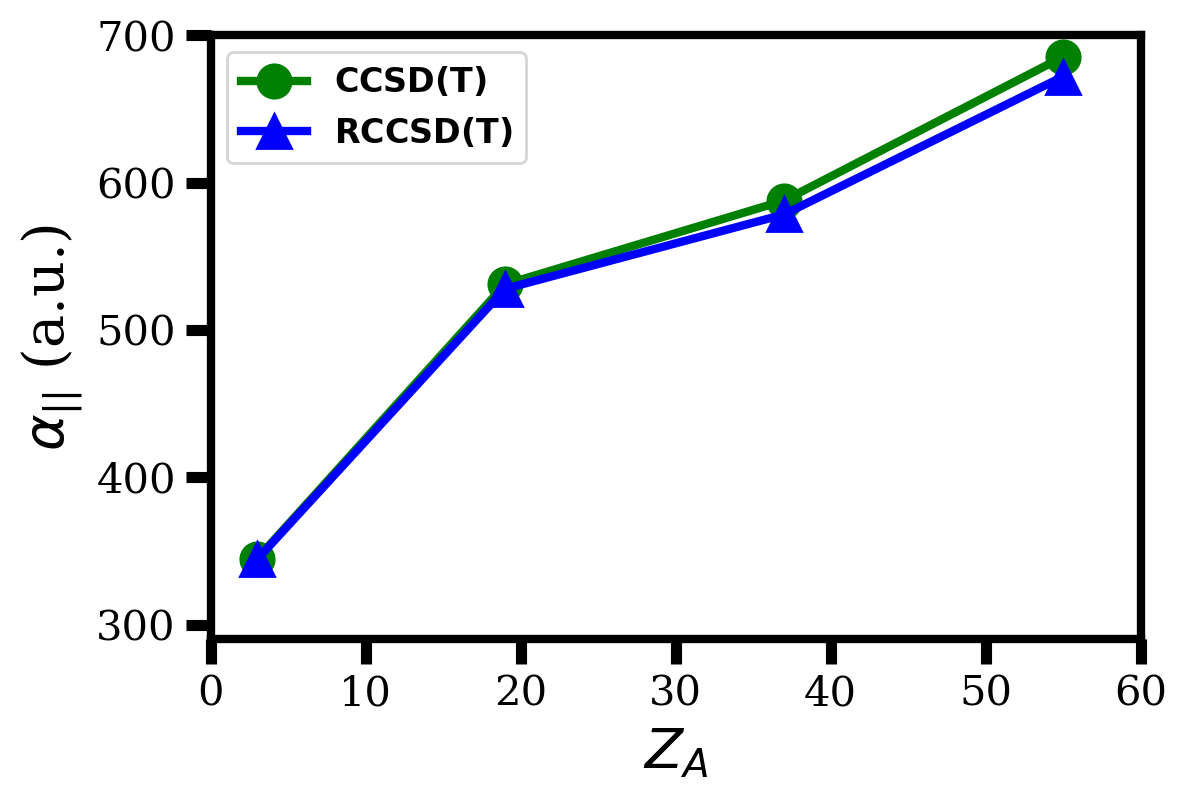} & \includegraphics[width=5cm, height=4.5cm]{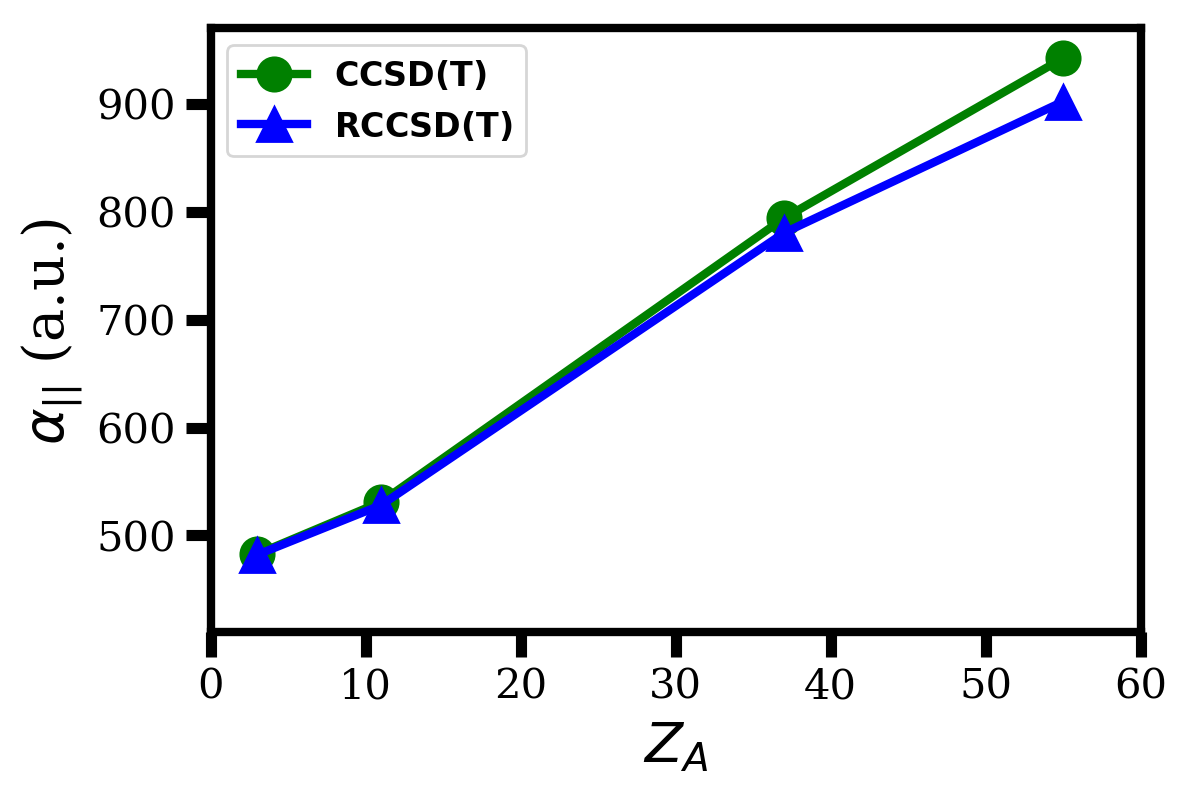} \\
(a) & (b) & (c) \\
\includegraphics[width=5cm, height=4.5cm]{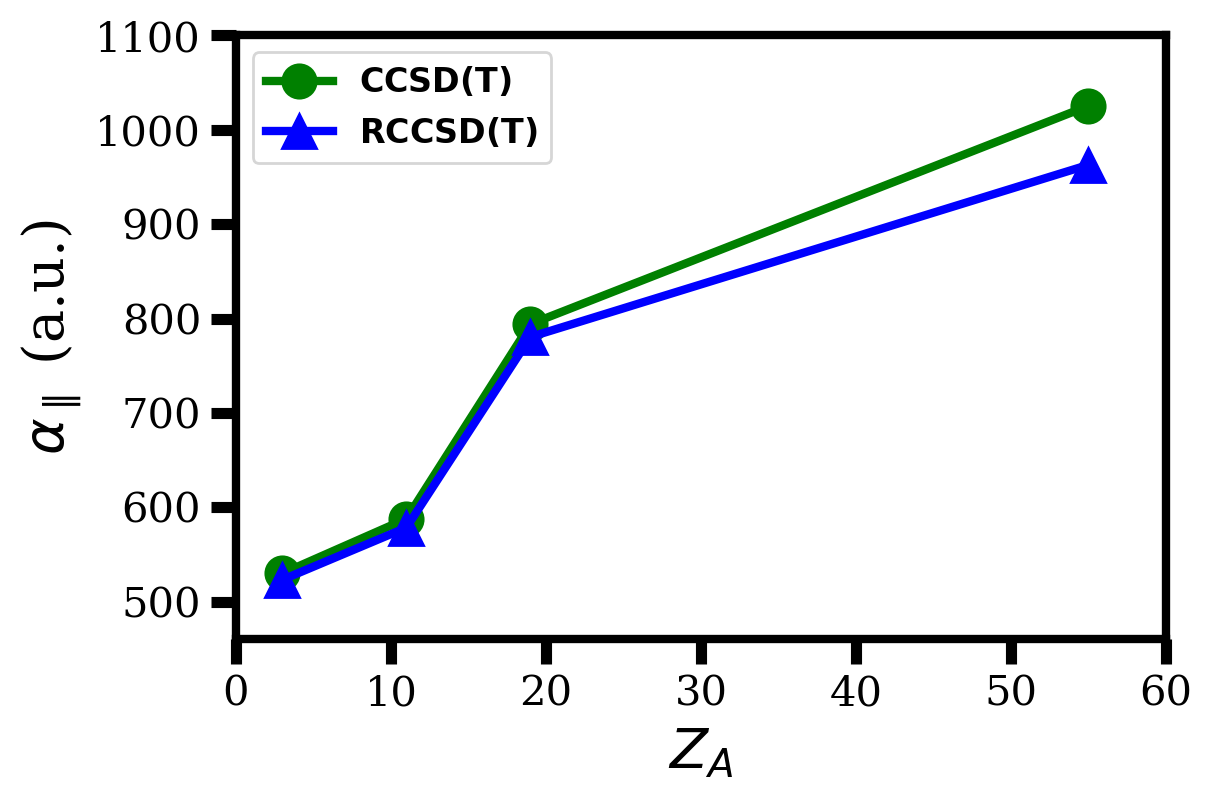} & \includegraphics[width=5cm, height=4.5cm]{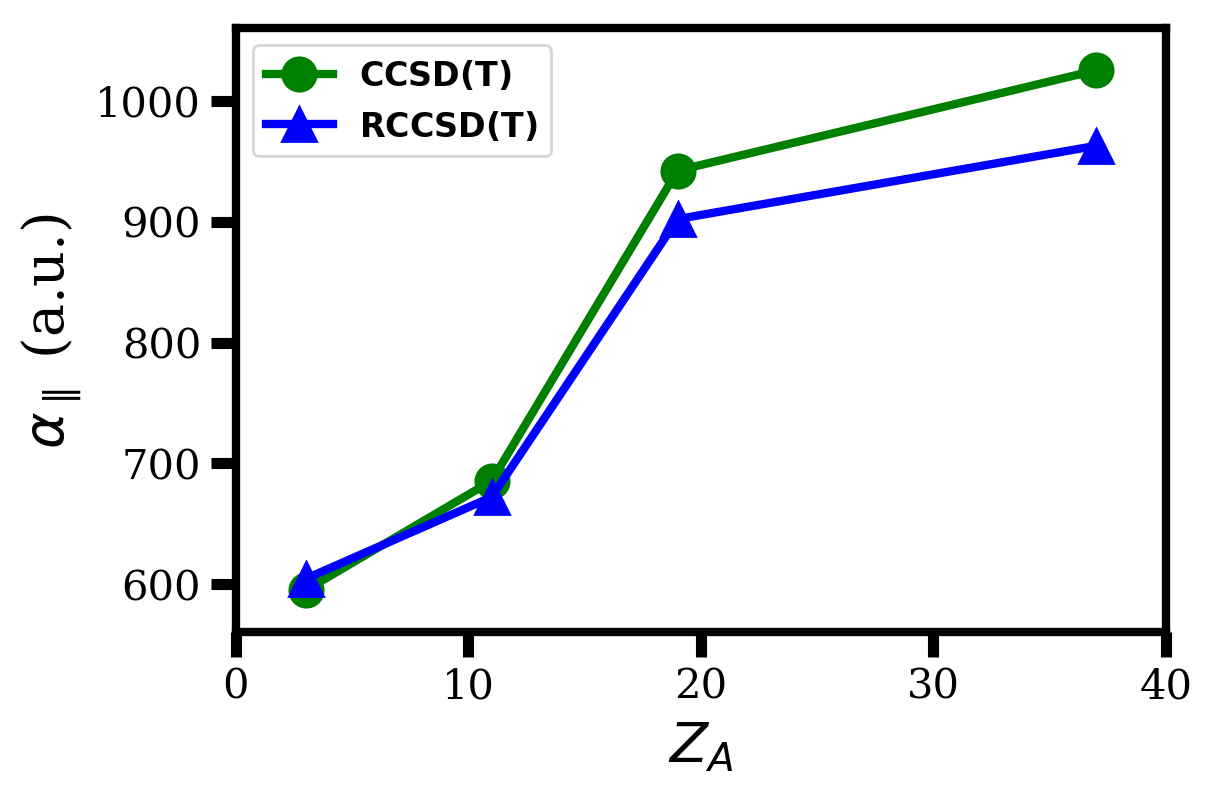} &\\
(d) & (e) & \\
\end{tabular}
\caption{(Colour online) Graphs illustrating the differences between the CCSD(T) and RCCSD(T) values of the parallel components of polarizability (in a.u.) of (a) Li-, (b) Na-, (c) K-, (d) Rb-, and (e) Cs-families of heteronuclear alkali dimers against the atomic number $Z_A$.}
\label{fig:figure3}
\end{figure*}

We will use this notation hereafter. Further, one defines polarizability anisotropy as the difference between the parallel and perpendicular components of the polarizability tensor, and is given by 
\begin{eqnarray}
\Delta\alpha &=& \alpha_{\parallel}-\alpha_{\perp}. \nonumber
\end{eqnarray}
To evaluate the energy, we need to take recourse to a quantum many-body theory. Among the various approximation methods, a very efficient one is the (R)CC method, due to its advantages over other existing ones~\cite{Bishop,Crawford}. The (R)CC method takes into account the electron correlation effects in terms of virtual excitations. The (R)CC wave function, $\arrowvert\Psi\rangle$, is given by 
\begin{equation}
\arrowvert\Psi\rangle=e^{T}\arrowvert\Phi_0\rangle,
\end{equation}
where $\arrowvert\Phi_0\rangle$ is the HF or DF wave function in the non-relativistic or relativistic calculations respectively, while $T$ is known as the excitation operator, which takes electrons from occupied orbitals to virtual ones in an N-electron system. It is given in second quantized form as  
\begin{equation}
T=T_1+T_2+T_3+\cdots+T_N. 
\end{equation}
Here, $T_i$ is the $i^{th}$ excitation operator. In the (R)CCSD approximation, the single and double excitation operators are given by 
\begin{eqnarray}
T_1&=&\sum_{i,a}t_{i}^aa^\dagger i,\\
\text{and} \ \ \ \ T_2&=& \frac{1}{4} \sum_{i,j,a,b}t_{ij}^{ab}a^\dagger b^\dagger ji,
\end{eqnarray}
respectively. Here, $a^{\dagger}$ and $i$ are the creation and annihilation operators corresponding to electron from a virtual orbital (particle) and an occupied orbital (hole), respectively, and $t_i^a$ is the amplitude of a single excitation for a given $i$ and $a$. For a double excitation, the corresponding amplitude is given by  $t_{ij}^{ab}$. 

The amplitudes of the (R)CC excitation operators are obtained by using the DIRAC16 program \cite{Dirac16}. The one- and two-body integrals are acquired by considering the electronic part of the non-relativistic molecular Hamiltonian, given by 
\begin{eqnarray}
H =  \sum_i \left [ \frac{p_i^2}{2} + V_{nuc}(r_i) \right ] + \frac{1}{2} \sum_{i,j} \frac{1}{r_{ij}} ,
\end{eqnarray}
and the 4-component Dirac-Coulomb (DC) Hamiltonian, given by 
\begin{eqnarray}
H &=&  \sum_i \Lambda_i^+ \left [ c\mbox{\boldmath$\alpha$}_i\cdot \bm{p}_i+\beta_i c^2 + V_{nuc}(r_i) \right ] \Lambda_i^+ \nonumber \\  && + \frac{1}{2} \sum_{i,j} \Lambda_i^+ \Lambda_j^+ \frac{1}{r_{ij}}  \Lambda_i^+ \Lambda_j^+ 
\end{eqnarray}
for the relativistic calculations. In the above expressions, $\bm{p}$ is the momentum operator. $V_{nuc}(r)$ is the nuclear potential, given by $\frac{Z_A}{|\vec{r}_{i} - \vec{R}_A|}$ for a point nucleus, with $Z_A$ denoting the atomic number of the $A^{th}$ nucleus and $\vec{r}_i$ and $\vec{R}_A$ the position vectors of the $i^{th}$ electron and $A^{th}$ nucleus with respect to the origin, respectively. In our work, we use a Gaussian charge distribution for the nucleus~\cite{GTO}. The term $\frac{1}{r_{ij}}=\frac{1}{|\vec{r}_i - \vec{r}_j|}$ is the two-body Coulomb interaction operator between the electrons located at $r_i$ and $r_j$. Also, $c$ is speed of light, $\mbox{\boldmath$\alpha$}$ and $\beta$ are the 4-component Dirac operators, and $\Lambda^+$ is the operator that projects the relativistic Hamiltonian onto the positive energies of Dirac sea~\cite{Brown,Mitt}. We chose the same bond lengths as in Refs.~\cite{Deiglmayr,Zuchowski} for the alkali dimers, and they are 5.4518 a.u. for LiNa, 6.268 a.u. for LiK, 6.5 a.u. for LiRb, 6.93 a.u. for LiCs, 6.61 a.u. for NaK, 6.88 a.u. for NaRb, 7.27 a.u. for NaCs, 7.688 a.u. for KRb, 8.095 a.u. for KCs, and 8.366 a.u. for RbCs. We used Dyall's triple zeta (TZ) basis sets~\cite{Dyallbasis} for heavier nuclei (K, Rb, and Cs) and for lighter elements (Li, and Na), we opted for augmented correlation-consistent polarized core valance triple zeta (aug-cc-pCVTZ) basis functions~\cite{ccpv}. 

\begin{table*}[t]
\centering
\caption{
The values of perpendicular components of dipole polarizability, $\alpha_{\perp}$ (in a.u.), both from the non-relativistic and relativistic methods. We have also added results that are obtained in the previous works for comparing with our calculations.}
    \label{tab:table3}
\begin{tabular*}{\textwidth}{lc@{\extracolsep{\fill}}ccccccccc}
\hline
Method &LiNa&LiK&LiRb&LiCs&NaK&NaRb&NaCs&KRb&KCs&RbCs\\
\hline \hline
\\
\multicolumn{11}{c}{\textbf{This work}}\\
HF &203.04&282.32&306.98&347.67&321.88&352.61&402.27&516.23&605.67&681.61\\
CCSD &187.67&249.42&264.58&294.69&284.17&307.90&346.94&424.52&489.13&541.42\\
CCSD(T) &186.98&247.27&266.80&293.37&280.06&303.41&343.51&410.83&473.24&519.76\\
\\
DF &202.67&280.92&301.20&335.69&319.57&344.86&385.81&501.11&571.87&628.52\\
RCCSD &187.33&248.31&262.50&287.97&282.08&300.90&333.43&412.41&463.20&505.76\\
RCCSD(T) &186.44&246.13&260.10&286.01&277.90&296.02&328.84&398.58&446.81&484.38\\
\\
\multicolumn{11}{c}{\textbf{Previous works}}\\
CCSD(T)\cite{Urban}&188.8&246.6&268.7&&268.7&303.2&&411.5&&\\
CCSD(T) \cite{2003} &187.7&&&&&&&&&\\
TDGI \cite{2003} &183.1&&&&&&&&&\\
CI\cite{Deiglmayr}&181.8&236.2&246.5&262.5&262.3&280.3&304.2&382.9&425.62&492.3\\
\hline \hline
    \end{tabular*}
\end{table*}

After obtaining the (R)CC amplitudes, the energy ($\Delta E$) is calculated by
\begin{eqnarray}
\Delta E=\langle\Phi_0\arrowvert H (1+T_1+T_2+\frac{1}{2}T_1^2)\arrowvert\Phi_0\rangle_C,\label{ccsdcorr}
\end{eqnarray}
where the subscript, `$C$', means that each term in the resulting expansion is fully contracted~\cite{Kvas}. 
We chose an external electric field perturbation, $\mathcal{E}$, with a strength of $0.0001$ a.u., for all our FF calculations. For heavier molecules, we cut-off electron excitations to high-lying virtuals to reduce the computational cost, as their contributions are negligible. For NaCs, a cut-off of $2000$ a.u. was imposed, while for the KRb, KCs and RbCs molecules, we cut-off all the orbitals possessing energies above $1000$ a.u..  We used a three-point central difference formula for our FF calculations of PDMs and static dipole polarizabilities. We have also systematically tested our numerical procedures by computing these properties using a five-point central difference scheme. 

\begin{figure*}[t]
\centering
\begin{tabular}{ccc}
\includegraphics[width=5cm, height=4.5cm]{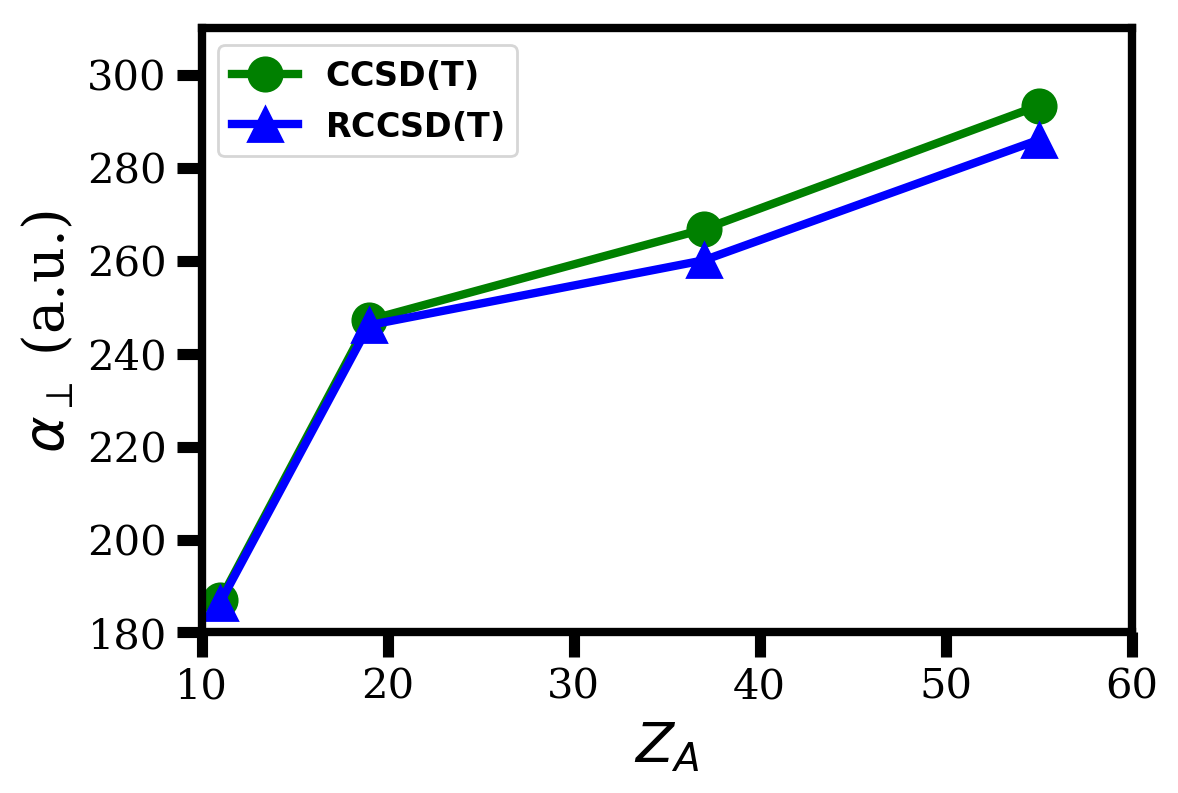} & \includegraphics[width=5cm, height=4.5cm]{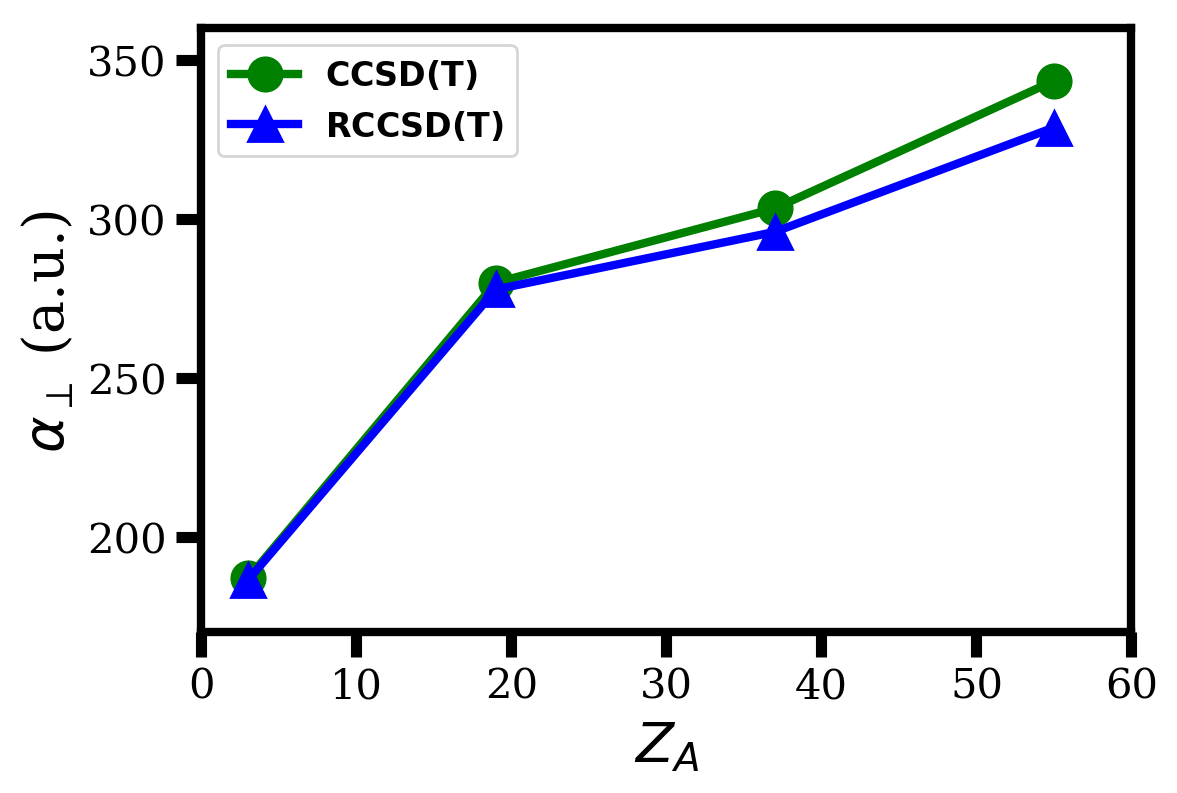} & \includegraphics[width=5cm, height=4.5cm]{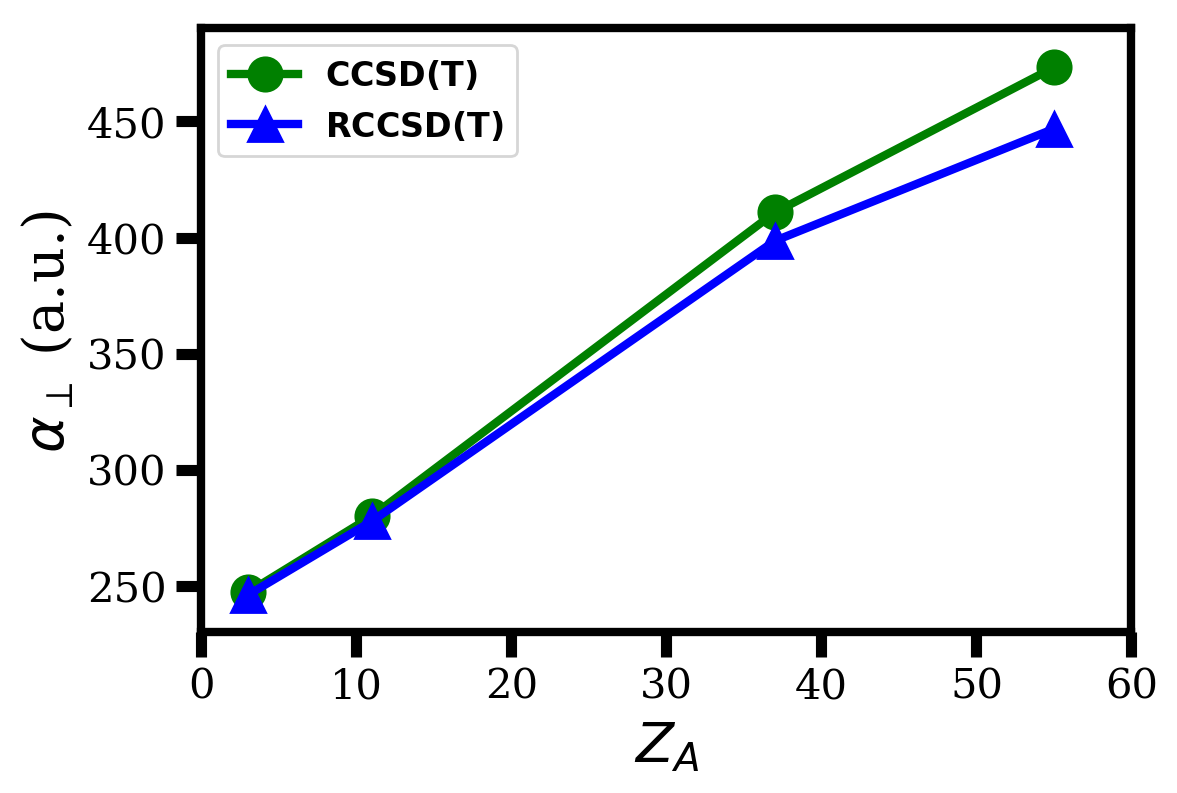} \\
 (a) & (b) & (c) \\
\includegraphics[width=5cm, height=4.5cm]{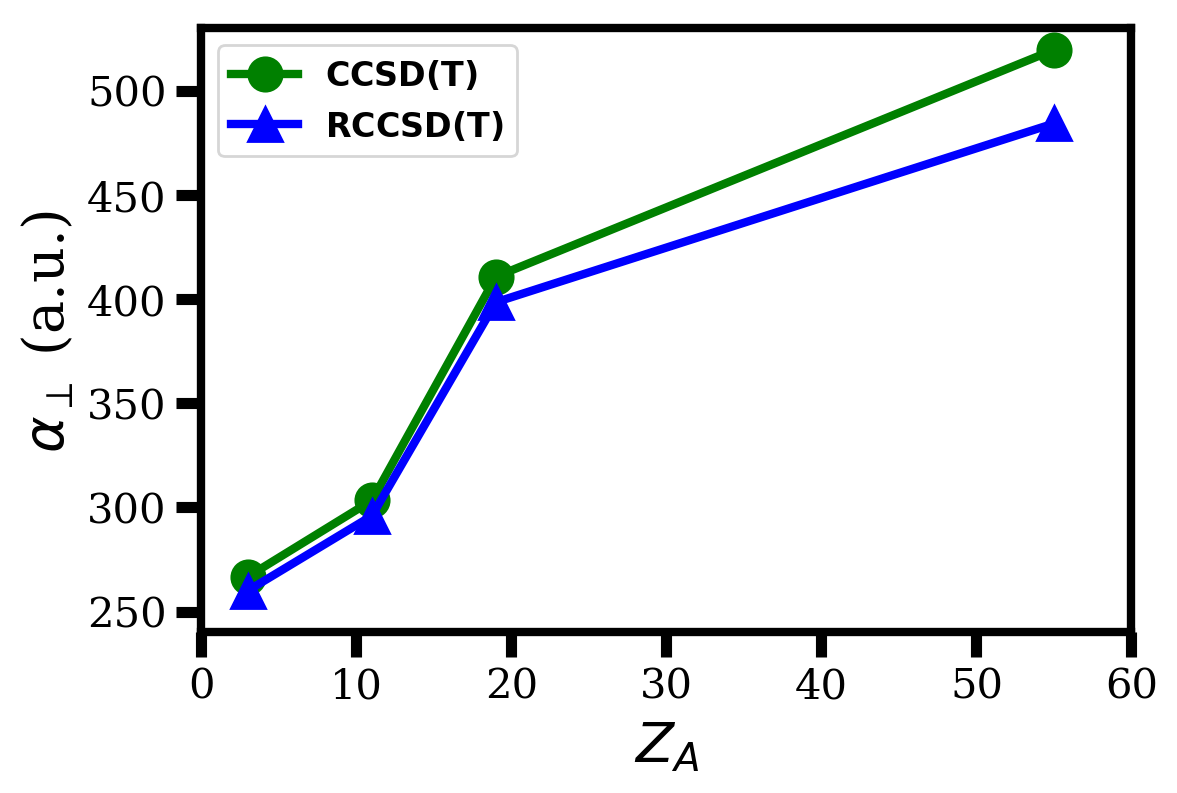} & \includegraphics[width=5cm, height=4.5cm]{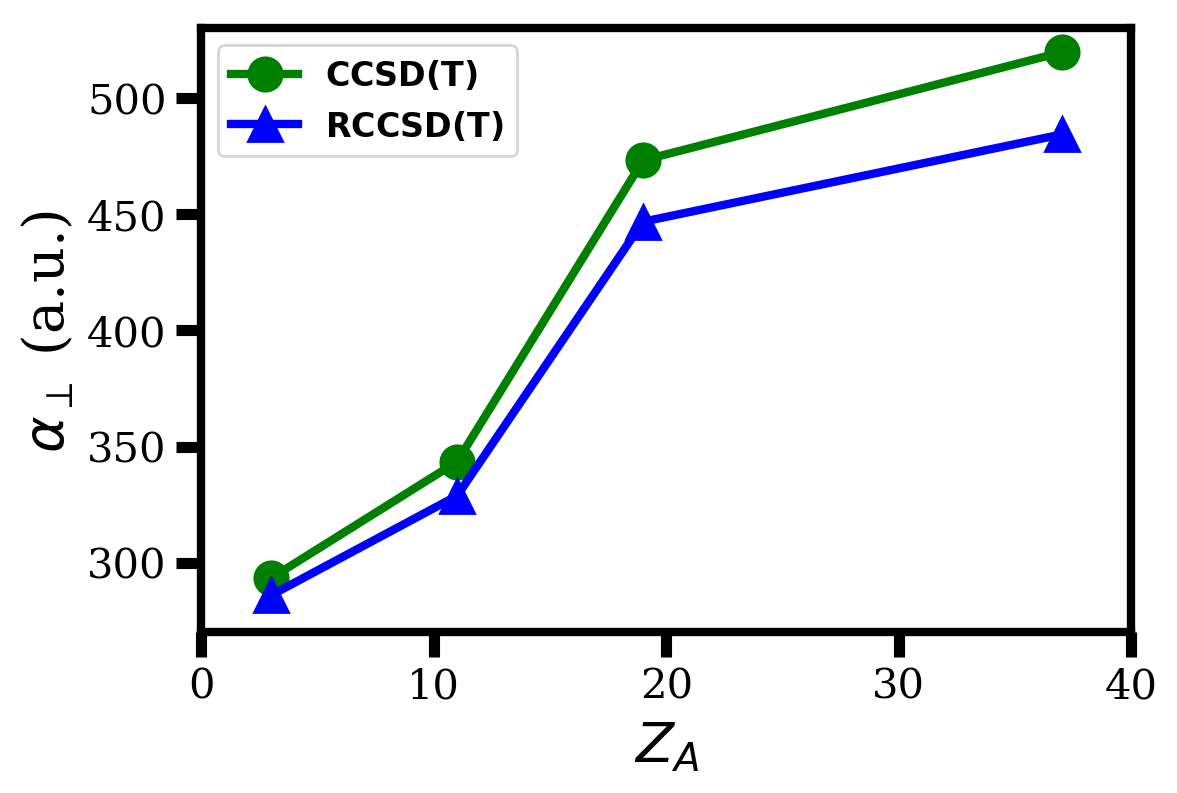} & \\
(d) & (e) \\
\end{tabular}
\caption{(Colour online) Graphs showing the departure between the values of perpendicular components of polarizability (in a.u.) obtained from the CCSD(T) and RCCSD(T) calculations of different families of heteronuclear alkali dimers. The sub-figures (a) shows the trends in the Li family, while (b), (c), (d), and (e) correspond to Na, K, Rb, and Cs families, respectively.}
\label{fig:figure4}
\end{figure*}

\section{Results and discussion} \label{sec3}

\begin{figure*}[t]
\centering
\begin{tabular}{ccc}
\includegraphics[width=5cm, height=4.5cm]{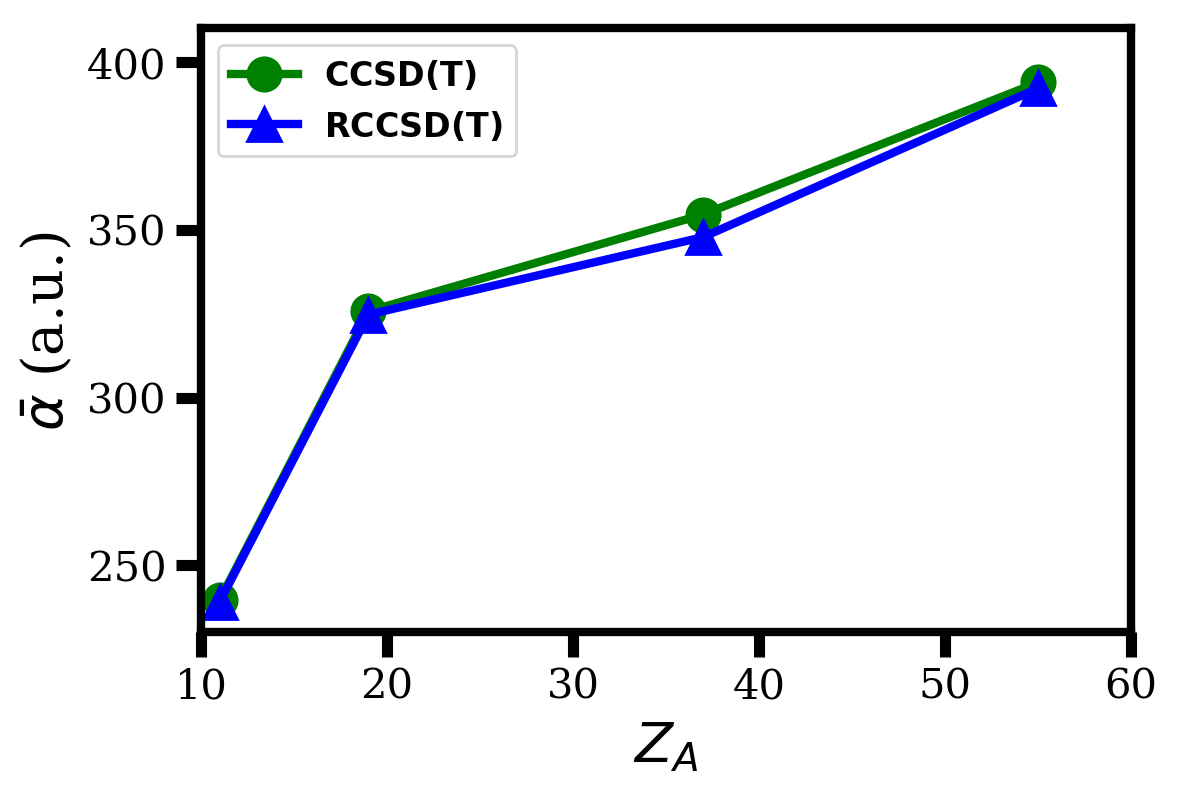} & \includegraphics[width=5cm, height=4.5cm]{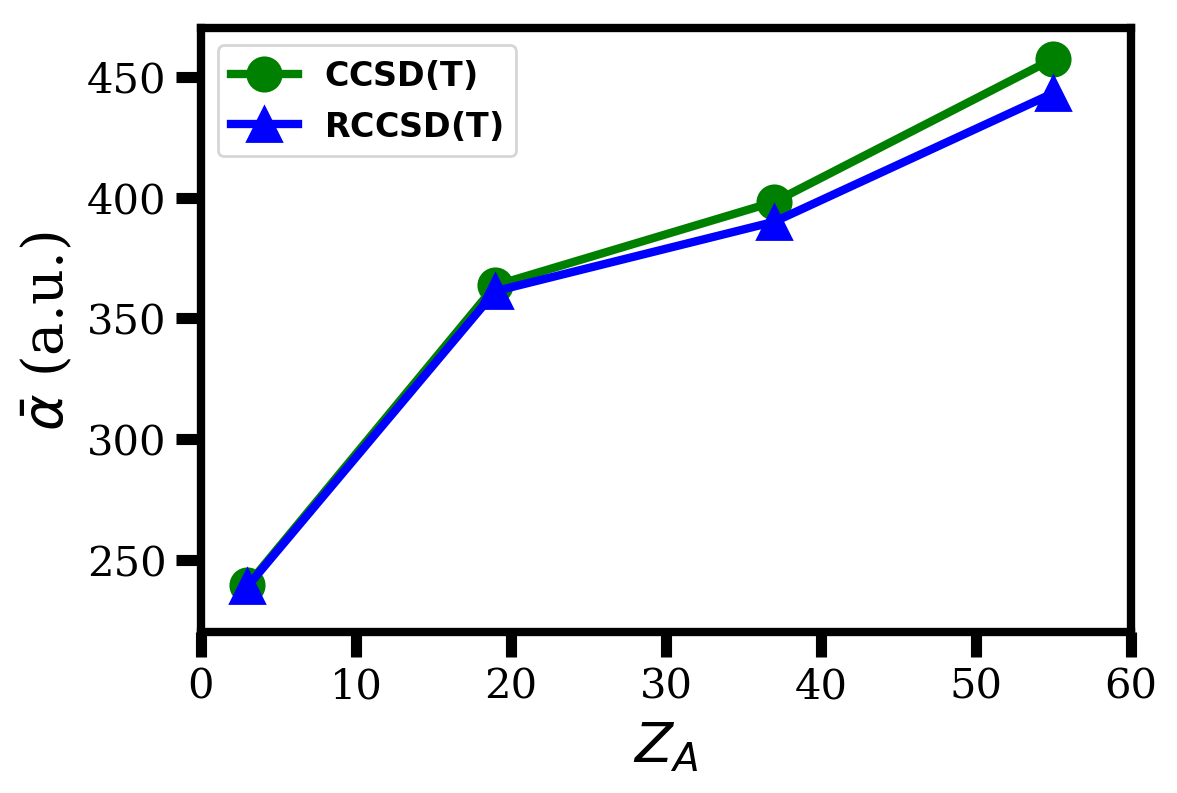} & \includegraphics[width=5cm, height=4.5cm]{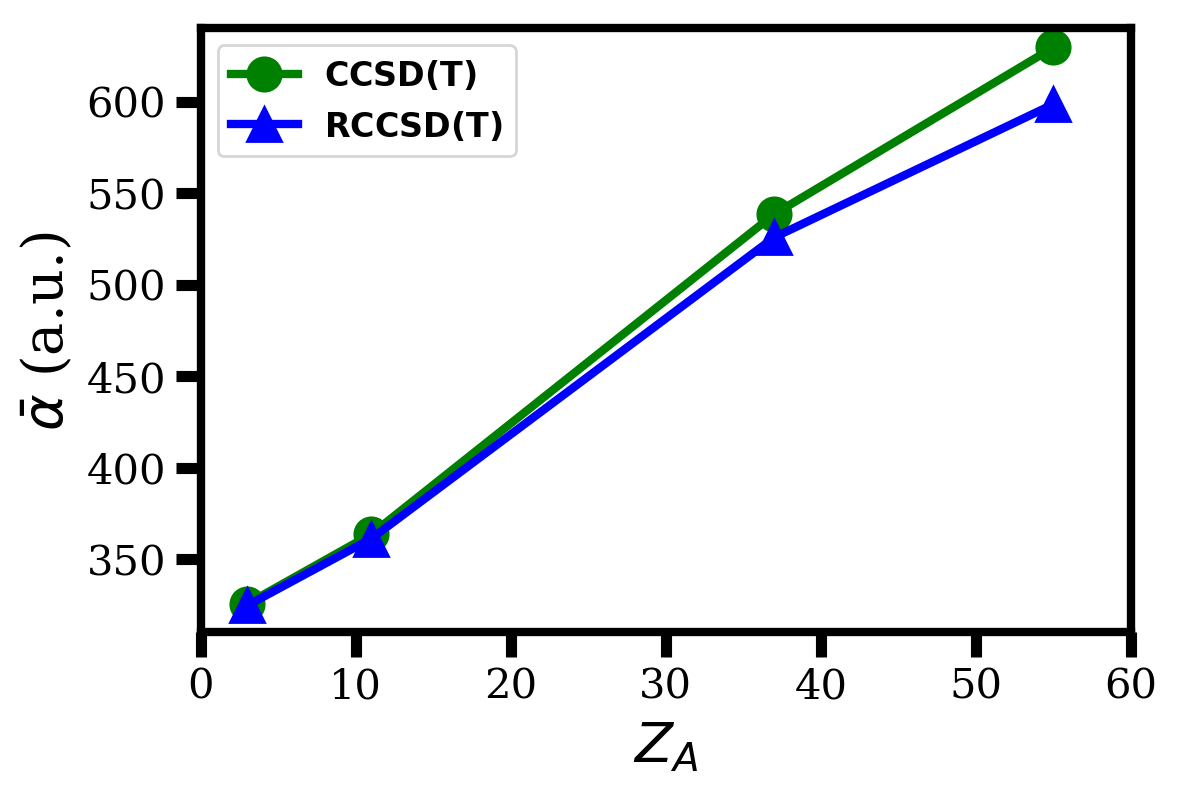}\\
(a) & (b) & (c) \\
\includegraphics[width=5cm, height=4.5cm]{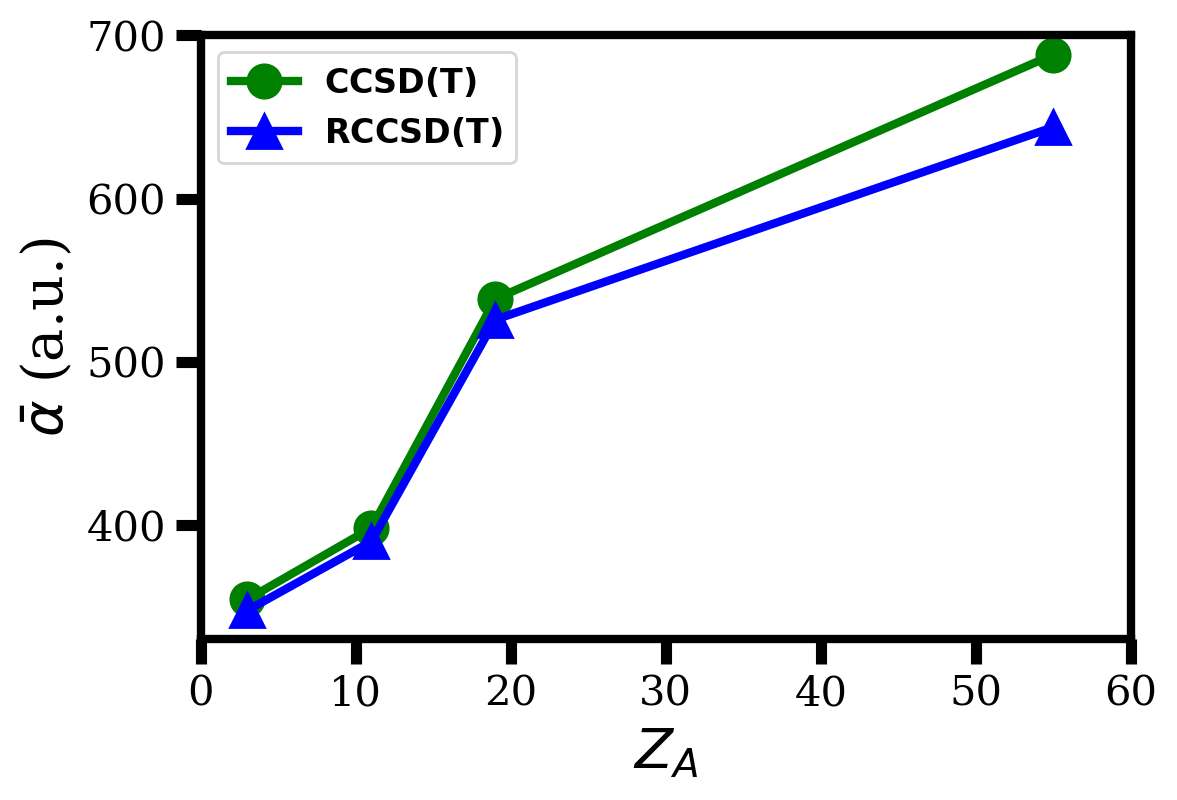} & \includegraphics[width=5cm, height=4.5cm]{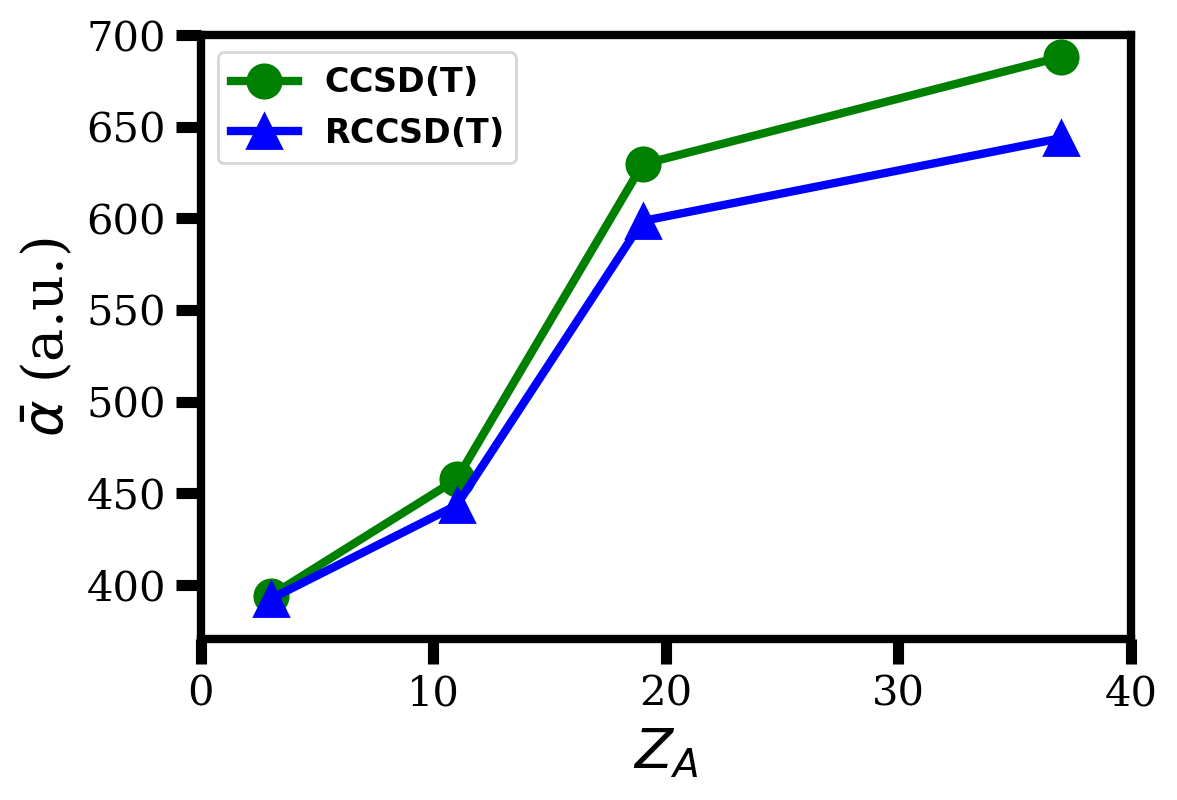} & \\
(d) & (e) & \\
\end{tabular}
\caption{(Colour online) Plots depicting the fork between the average dipole polarizabilities (in a.u.) in the (a) Li-, (b) Na-, (c) K-, (d) Rb-, and Cs-families from the CCSD(T) and RCCSD(T) methods.}
\label{fig:figure5}
\end{figure*}

In this section, we examine our results for the PDMs, and then polarizabilities, followed by error analysis. We discuss in detail the trends observed for the PDMs, based on Table \ref{tab:table1} and Fig. \ref{fig:figure1}. We then proceed to compare our results with previous works. Fig. \ref{fig:figure2} presents our accurately calculated relativistic $\mu$ values for LiNa, obtained at complete basis set (CBS) limit, and its excellent agreement with experiment. In Tables \ref{tab:table2} and \ref{tab:table3}, we present our results for $\alpha_{\parallel}$ and $\alpha_{\perp}$, respectively, which are complemented by Figs. \ref{fig:figure3} and \ref{fig:figure4}. This is followed by discussions on the average polarizabilities and anisotropies, with the corresponding data presented in Table \ref{tab:table4} and Fig. \ref{fig:figure5}, and Table \ref{tab:table5} and Fig. \ref{fig:figure6}, respectively. Table \ref{tab:table6}, and the text accompanying it, illustrates the importance of relativistic calculations for the isotropic Van der Waals $C_6$ coefficients in molecule-molecule interactions, while Fig. \ref{fig:figure7} shows the linear variation of the components of polarizabilities with volume. We then present detailed error estimates, one of which is shown explicitly in Table \ref{tab:table7}. 

\begin{table*}[t]
\centering
    \caption{The average values of dipole polarizability, $\bar{\alpha}$ (in a.u.), of the alkali dimers from both our and previous calculations. We have also given experimental values for the comparison.}
    \label{tab:table4}
\begin{tabular*}{\textwidth}{lc@{\extracolsep{\fill}}ccccccccc}
\hline \hline
Method&LiNa&LiK&LiRb&LiCs&NaK&NaRb&NaCs&KRb&KCs&RbCs\\
\hline
\\
\multicolumn{11}{c}{\textbf{This work}}\\
HF &236.02&330.19&361.64&407.03&381.97&419.72&476.65&616.26&722.40&815.93\\
CCSD &238.79&323.10&347.35&384.41&363.53&396.64&449.10&547.34&635.84&702.25\\
CCSD(T) &239.58&325.72&354.55&393.90&363.71&398.24&457.50&538.59&629.62&688.29\\
\\
DF &235.64&329.12&355.93&400.29&379.78&413.23&464.46&603.72&692.06&763.98\\
RCCSD &238.40&322.23&344.27&384.39&361.30&389.75&438.37&536.04&608.87&660.90\\
RCCSD(T) &239.06&324.73&347.82&392.14&361.27&390.01&443.27&525.88&598.61&643.87\\
\\
\multicolumn{11}{c}{\textbf{Previous works}}\\
\textbf{Experiment} &&&&&&&&&&\\
Ref.~\cite{Graff}&269.93(33.74)&&&&&&&&&\\
Ref.~\cite{Tarnovsky}&&&&&344.16(26.99)&&&&600.60(42.24)\\
Ref.~\cite{Tarnovsky}*&249.69&377.91&384.65&465.63&391.40&398.15&479.13&526.37&607.35&614.10\\
\\
\multicolumn{1}{l}{\textbf{Theory}} &&&&&&&&&&\\
CCSD(T)~\cite{Urban}&243.23&326.00&365.20& &365.57&404.23& &555.13& &\\
CI~\cite{Deiglmayr}&237.0&320.7&339.1&374.0&351.3&377.5&426.4&504.8&571.1&602.8\\
CCSD(T)~\cite{Zuchowski}&237.7&324.2&347.2&391.9&358.1&387.1&439.3&523.5&596.0&638.6\\
\hline \hline
    \end{tabular*}
    \\
    *These values are not strictly experimental, as they are obtained by combining measured homonuclear polarizability with an empirical rule. The rule may not always hold, as evident from the difference in their results that they arrived at by using this approach as compared to their experimental value, for NaK. 
\end{table*}

Below, we discuss our results on PDMs and polarizabilities of the considered molecules along with their trends that we observe from our calculations. We then proceed to compare our values with the available ones from literature, for each property. While reporting the trends, we do so within a family (for example, Li family refers to LiA; A=Na, K, Rb, and Cs) and between them, rather than look for trends by arranging the molecules in the increasing order of the number of electrons. This is because two molecules that are next to each other in the number of electrons could be very different, as one may possess a combination of light-heavy nuclei and other moderate-moderate nuclei. We will see in the subsequent paragraphs that ordering molecules in this manner, i.e., by family, provides better insights in trends. A useful quantity to define for the following discussions is the percentage fraction change, $F$, given by the magnitude of $(\frac{Rel-NR}{Rel}\times 100)$ for a property; with `$Rel$' and `$NR$' being the respective relativistic and non-relativistic values from a given approach. Basically, $F$ quantifies the corrections due to the relativistic effects in a molecule, for that property. 

\subsection{Results for PDMs}

Here, we analyze the trends in the PDM (whose results are provided in Table~\ref{tab:table1}), starting with the Li family. In this paragraph, we will first examine the effects of correlations on the PDMs including the roles of partial triples, followed by a detailed report on relativistic effects. We will adopt this order of discussing results for the polarizabilities too. By comparing the three methods that we have employed, we observe that for a given molecule in a family, the inclusion of correlation effects steadily decreases the value of $\mu$, in both the non-relativistic and relativistic cases. We observe that partial triples in the (R)CCSD(T) methods reduce the values of PDM as compared to those from the (R)CCSD methods. This effect could be as large as 17 percent, as in the case of KRb. We now move on to the roles of relativistic effects. A visual inspection from Fig. \ref{fig:figure1}(a) shows that for the Li family, the gap between the non-relativistic and relativistic results widens as a molecule becomes heavier. When we inspect more closely and calculate $F$, we observe that it indeed increases, but from LiK through LiCs. LiNa displays  more change in its PDM with the inclusion of relativistic effects than LiK, in all the three methods (in the HF, CCSD and CCSD(T) methods as well as in the DF, RCCSD and RCCSD(T) methods). We also observe that relativity increases the PDM of LiCs at the RCCSD(T) level of theory by about 18 percent, which is clearly not negligible. The trends in the Na family (see Fig. \ref{fig:figure1}(b)) are qualitatively similar to those in the Li family. Again, with the exception of LiNa, we observe a monotonic increase in $F$, with relativistic effects accounting for as high as about 21 percent for NaCs. In the K family, whose trend is shown in Fig. \ref{fig:figure1}(c), we observe the first deviation from monotonic behaviour as the PDMs decrease from KLi to KRb, and then increase from KRb to KCs. We observe  similar trends with the Rb family as well (see Fig. \ref{fig:figure1}(d)). In the Cs family (see Fig. \ref{fig:figure1}(e)), we report a monotonic decrease in the values of PDM. Also, we see that the relativistic effects play significant roles starting from the K family, with $F$ being about 50 percent for KRb, KCs, and RbCs. In light of the significance of relativistic effects for these systems, our RCCSD(T) calculations for the heteronuclear alkali dimer molecules are the most accurate, to the best of our knowledge. Lastly, we comment on the importance of triple excitations, at the CCSD(T) and RCCSD(T) levels. We would like to comment on some of the recent works on the PDMs of alkali dimers, and compare their results with ours below. 

\subsubsection{LiNa}\label{linap}

There are a number of calculations on the PDM of LiNa; for example, see Refs.~\cite{S. Green,P. Rosmus,Peter Habitz,Ron Shepard,Muller,Igel Mann}. Most of these earlier works were carried out by employing non-relativistic methods, and some of the results were at odds with the experimental values. We focus on and compare here our results with experiments, and the more recent theoretical studies. 

Dagdigian \textit{et al}~\cite{P. J. Dagdigian}, in 1971, reported the PDM of LiNa to be 0.18(1) a.u.. In the experiment, the measured quantity is actually $\mu^2/B$, where $B$ is the rotational constant of the molecule. Then, $B$ was obtained by using Badger's rule, which required the knowledge of the then-existing literature values for the spectroscopic constants of the molecule. In a subsequent work~\cite{Dag2}, they improved their value for $B$, and obtained a PDM of 0.1822(7) a.u., with much lesser uncertainty. A third work from the same group~\cite{Graff} found the quantity to be 0.1822(8) a.u., by performing a molecular beam resonance experiment. A fourth experimental result was obtained in 1982, as 0.1777(2) a.u.~\cite{F. Engelke}, using laser-induced fluorescence spectra. They too obtained $B$, and hence the PDM. A PDM of 0.18 a.u. was reported by Tarnowsky \textit{et al}~\cite{Tarnovsky}. They estimated the property from the empirically derived formula  $\mu_{XY}=C(\bar{\alpha}_{X_2}-\bar{\alpha}_{Y_2})$, where $\mu_{XY}$ corresponds to the PDM of a molecule made of atoms $X$ and $Y$, and $\alpha_{X_2}$ corresponds to the average polarizability of a homonuclear dimer of type $X_2$, and likewise for $\alpha_{Y_2}$. The PDM, $\mu_{XY}$, was computed by the authors with a fitting procedure, which in turn required their measured homonuclear polarizabilities of $\mathrm{Li_2}$ and $\mathrm{Na_2}$ as well as their PDMs taken from the then-recent literature. 

As a survey of literature described above shows that five different measurements give almost the same value, the experimental result of about 0.18 a.u. itself is very reliable. However, we note that there is a strong tension in results between experiment and theory, as seen from Table~\ref{tab:table1}. In fact, for a specified method employed by a work on alkali dimers, the agreement between experiment and theory is the least for LiNa among other reported alkali dimers. One such example is Ref.~\cite{Igel Mann}, where the authors employ configuration interaction (CI) method to find this issue. We proceed now with discussion of the results obtained from more recent calculations. The work by Urban and Sadlej~\cite{Urban} considered the electron correlation contribution due to the next-to-valence electrons of the two atoms forming a dimer, and reported a PDM of 0.17 a.u.. They employed the CCSD(T) method, and added relativistic corrections due to mass-velocity and Darwin (MVD) terms. A subsequent work~\cite{1999} employed CI in the singles and doubles approximation considering correlations only among ten electrons to obtain 0.19 a.u.. Aymar and Dulieu~\cite{M. Aymar} had employed a full valence CI approach with pseudo-potentials (PP), and included relativistic effects via MVD terms. They considered three different basis sets, which we denote in the Table~\ref{tab:table1} as Basis A, B, and C. They obtained 0.221 a.u. and 0.218 a.u., by using basis sets A and B, respectively. Vanner \textit{et al}~\cite{Vanner} obtained 0.228 a.u. using their CI approach involving the  perturbation of a multi-configuration wave function selected iteratively, in a PP approach. Core-polarization and core-valence interactions were partially considered by using $l$-dependent core-polarization potentials. Zuchowski \textit{et al}~\cite{Zuchowski} computed the PDM of LiNa to be 0.19 a.u., using the CCSD(T) method, and employed a cc-pCV5z basis, augmented with diffuse functions. A PDM of 0.21 a.u. was reported by Federov \textit{et al}~\cite{Dmitry A. Fedorov}, and was calculated by using CC in the singles, doubles and tripes approximation (CCSDT) with an aug-cc-pV5Z basis. We obtain $\mu$=0.22 a.u., using the RCCSD(T) approach, with a TZ quality basis that includes diffuse functions, to capture far-nuclear region information, which can be important for the evaluation of PDM. Our relativistic TZ result, obtained without imposing any cut-off on virtuals, and with the aug-cc-pCVTZ basis for Li and Na, is consistent with that obtained by the non-relativistic benchmark CCSDT calculation in a 5Z basis from Ref.~\cite{Dmitry A. Fedorov}. However, we obtain similar results for different reasons, as our work is relativistic and uses diffuse functions in the basis, while theirs employs neither, but uses a higher quality 5Z basis. We observe from our results in Table~\ref{tab:table1} that relativistic effects in our RCCSD(T) calculations reduce $\mu$ by about 4.5 percent. This indicates that the inclusion of relativistic effects may reduce the dipole moment obtained in Ref.~\cite{Dmitry A. Fedorov} towards the experimental result. Also, the table shows that the work in Ref.~\cite{Dmitry A. Fedorov} over-estimates the PDM value, possibly due to triple excitations displaying a tendency to increase its value, besides not accounting for relativistic effects. Our TZ result includes relativistic effects, but over-estimates the PDM, most likely due to our basis being limited to TZ quality. This can be seen from Ref.~\cite{Dmitry A. Fedorov}, where one observes that the PDM reduces with better quality basis, as shown in Table~\ref{tab:table1}. RCCSD(T) computations on LiNa is still feasible with a QZ basis. Therefore, we further improve our result for LiNa by using a quadruple-zeta (QZ) basis with diffuse functions added to them, and subsequently extrapolating to the CBS limit. We employ the two-point scheme by Helgaker~\cite{cbs0,cbs1} for CBS extrapolation, which is known to be simple and accurate. Fig.~\ref{fig:figure2} shows our results, and we obtain a value that is very close to experiment, at 0.178 a.u.. We have also verified that including Gaunt interaction does not alter the PDM, at the TZ and QZ levels of basis functions. 

\begin{table*}[t]
\centering
    \caption{The non-relativistic and relativistic values of dipole polarizability anisotropy, $\bar{\alpha}$ (in a.u.), reported at different levels of (R)CC theory and other methods. A list of previous works are added to the table for comparison with our results. } 
    \label{tab:table5}
\begin{tabular*}{\textwidth}{lc@{\extracolsep{\fill}}ccccccccc}
\hline \hline
Method&LiNa&LiK&LiRb&LiCs&NaK&NaRb&NaCs&KRb&KCs&RbCs\\
\hline
\\
\multicolumn{11}{c}{\textbf{This work}}\\
HF &98.94&143.60&163.97&178.07&180.27&201.05&223.15&300.10&350.20&402.96\\
CCSD &153.35&221.05&248.30&269.17&238.07&266.22&306.47&368.48&440.13&482.48\\
CCSD(T) &157.80&235.37&263.24&301.57&250.95&284.49&341.98&383.29&469.15&505.59\\
\\
DF &98.96&144.60&164.18&193.79&180.63&205.11&235.94&307.84&360.58&406.37\\
RCCSD &153.22&221.75&245.30&289.26&237.65&266.56&314.81&370.89&437.01&465.41\\
RCCSD(T) &157.85&235.81&263.16&318.39&250.11&282.48&343.28&381.90&455.41&478.47\\
\\
\multicolumn{11}{c}{\textbf{Previous works}}\\
\textbf{Experiment} &&&&&&&&&&\\
Ref.~\cite{Graff}&161.96(13.5)&&&&&&&&&\\
\\
\textbf{Theory} &&&&&&&&&&\\
CCSD(T)~\cite{Urban}&163.3&238.2&289.5& &2579&303.1& &430.9& &\\
CCSD(T)~\cite{Zuchowski}&156.3&234.5&262.0&317.8&247.2&279.2&339.4&367.6&436.1&462.1\\
CI~\cite{Deiglmayr}&165.8&253.5&277.8&334.5&266.9&291.7&366.5&365.8&436.7&491.7\\
\hline \hline
    \end{tabular*}
\end{table*}

\subsubsection{LiX; X = K, Rb, and Cs}

Our results for LiK and LiRb are in excellent agreement with experimental results from Dagdigian \textit{et al}~\cite{P. J. Dagdigian}, with our obtained values being well within the error bars. However, the other experimental result presented in  Ref.~\cite{F. Engelek} that is available for LiK provides a slightly higher value. Since Ref.~\cite{F. Engelek} uses an improved value for $B$ as compared to that used in  Ref.~\cite{P. J. Dagdigian}, we expect the former to be more accurate. We anticipate that calculations with an even higher quality basis than ours could account for this gap of about 1.5 percent between our work and experiment. For the heavier LiRb and LiCs molecules, even though our calculations and the existing theoretical works agree closely, we expect that our results must be an improvement over the existing theoretical works (we did not find any experimental result for LiCs for comparison), since relativistic effects begin to play an important role in these systems. The most recent calculation by Fedorov \textit{et al}~\cite{Dmitry A. Fedorov} employs a higher quality 5Z basis for the lighter Li as compared to our TZ basis, but for the more important heavier systems, our basis is better than their small-core relativistic effective core potentials (ECP). Moreover, they correlate all of their 9 explicit electrons (one valence electron from the outermost s orbital, and 8 from the next sp-shell) of K, Rb, and Cs in their work. We do not make any such approximations, and we correlate all electrons besides not cutting-off any virtuals in our RCCSD(T) calculations with a TZ basis for the $LiX$ molecules. The importance of relativistic effects is especially evident from the difference between our non-relativistic and relativistic results for LiCs. 

\subsubsection{The Na family}

Tarnovsky \textit{et al}~\cite{Tarnovsky} reported a PDM of 1.34 a.u. for NaK using an approach that combines measurement with an empirical rule, as discussed under Sec. \ref{linap}. The work in Ref.~\cite{Dag2} found the PDM to be 1.09(4) a.u., using their $B$ value, which in turn was obtained from an extrapolation of Badger's rule. Our result is within 2 percent of both the experimental value as well as the most recent theoretical work~\cite{Dmitry A. Fedorov}. The experimental values for NaRb and NaCs were obtained too with their respective $B$ values computed using an extrapolation of Badger's rule~\cite{Dag2}. Our results agree well with both experiment and recent calculations from other groups~\cite{Dmitry A. Fedorov,Zuchowski}. 

\subsubsection{The K and Rb families}

The last three molecules that we consider, viz. KRb, KCs, and RbCs, are made solely of relatively heavier atoms. Experimental values exist for KRb and RbCs, and our PDMs differ from the experimental results by about 8 percent and 4 percent, respectively. At this point, it is worth noting that the most recent non-relativistic calculation on the heavy KRb system using relativistic ECP, done in Ref.~\cite{Dmitry A. Fedorov}, differs from experiment by about 17 percent, indicating the importance of relativistic effects. We also observe that PDMs reported in Ref.~\cite{Zuchowski} agree very well with our TZ results (within 2 percent), with the exception of LiNa, where the difference is about 16 percent. They have employed ECPs with tailored valence basis sets for the heavier atoms. Although there is no indication from their work that they have performed relativistic calculations, their agreement with our results strongly suggests that they may have taken into account relativistic corrections in some form over their non-relativistic results; which is unclear to us from their paper. 

\begin{figure*}[t]
\centering
\begin{tabular}{ccc}
\includegraphics[width=5cm, height=4.5cm]{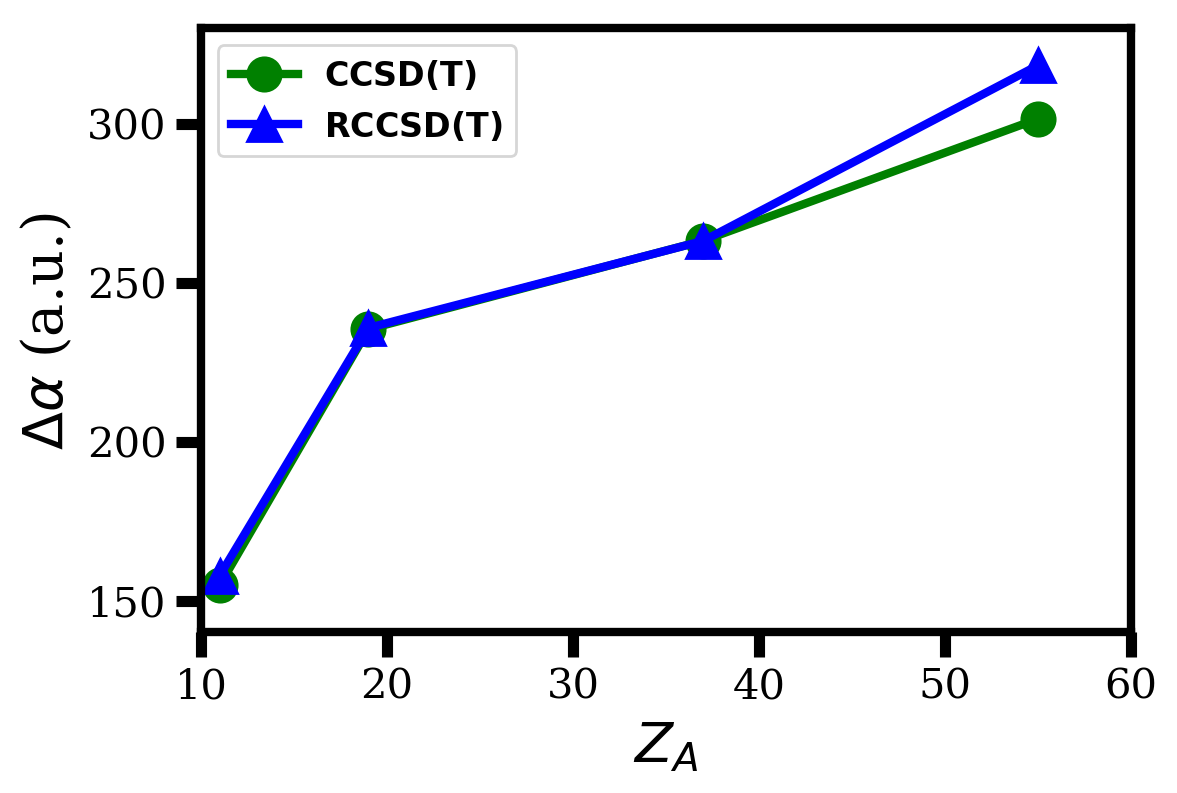} & \includegraphics[width=5cm, height=4.5cm]{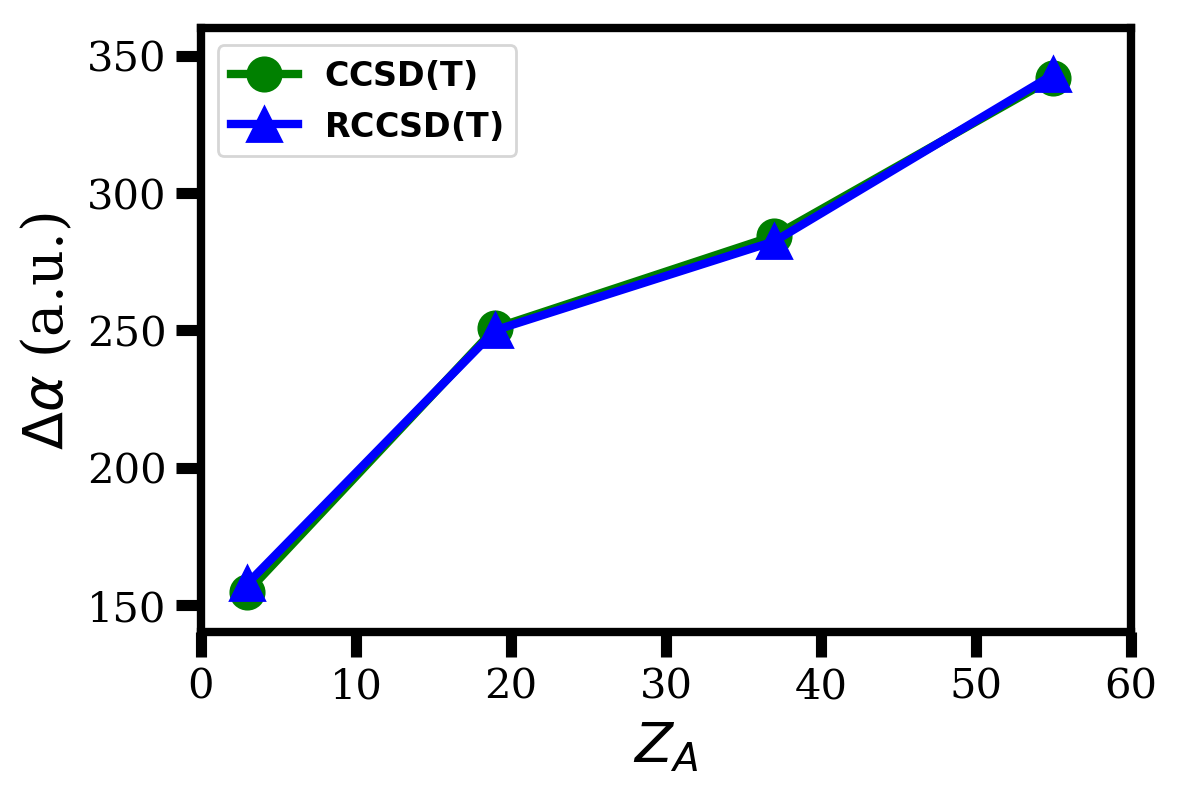} & \includegraphics[width=5cm, height=4.5cm]{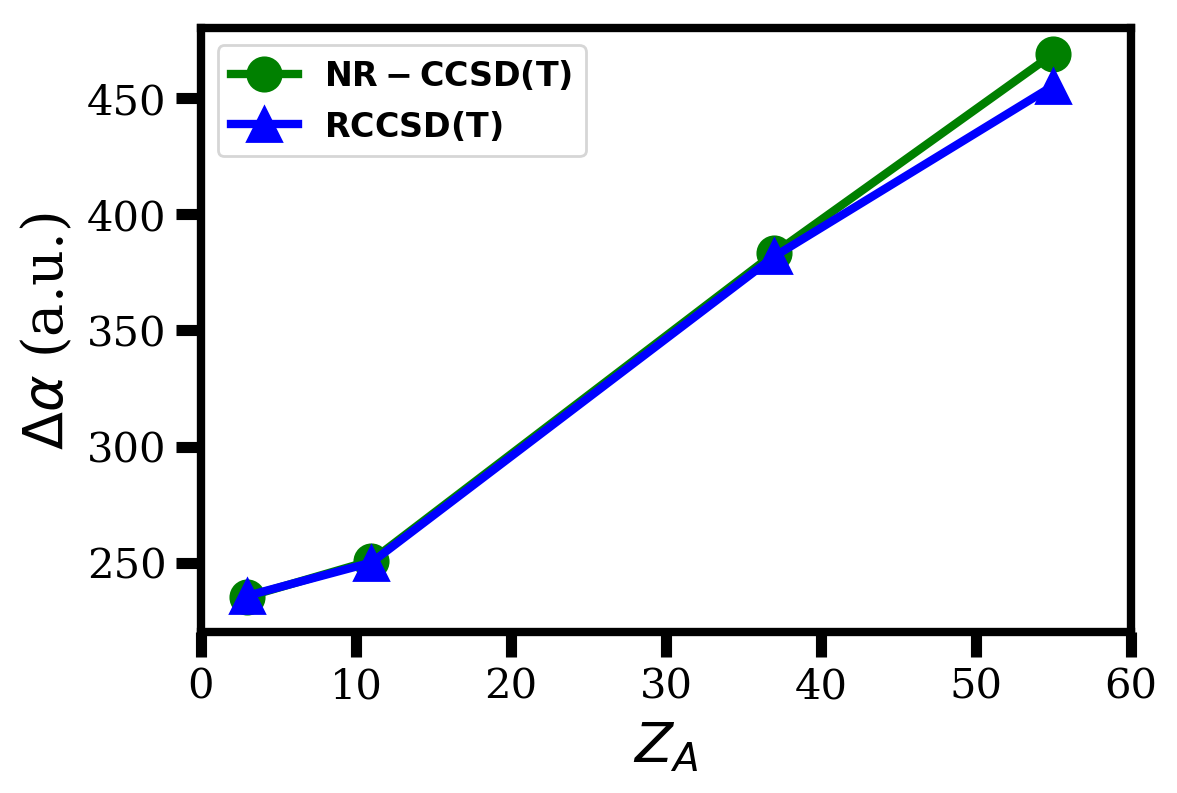}\\
(a) & (b) & (c) \\
\includegraphics[width=5cm, height=4.5cm]{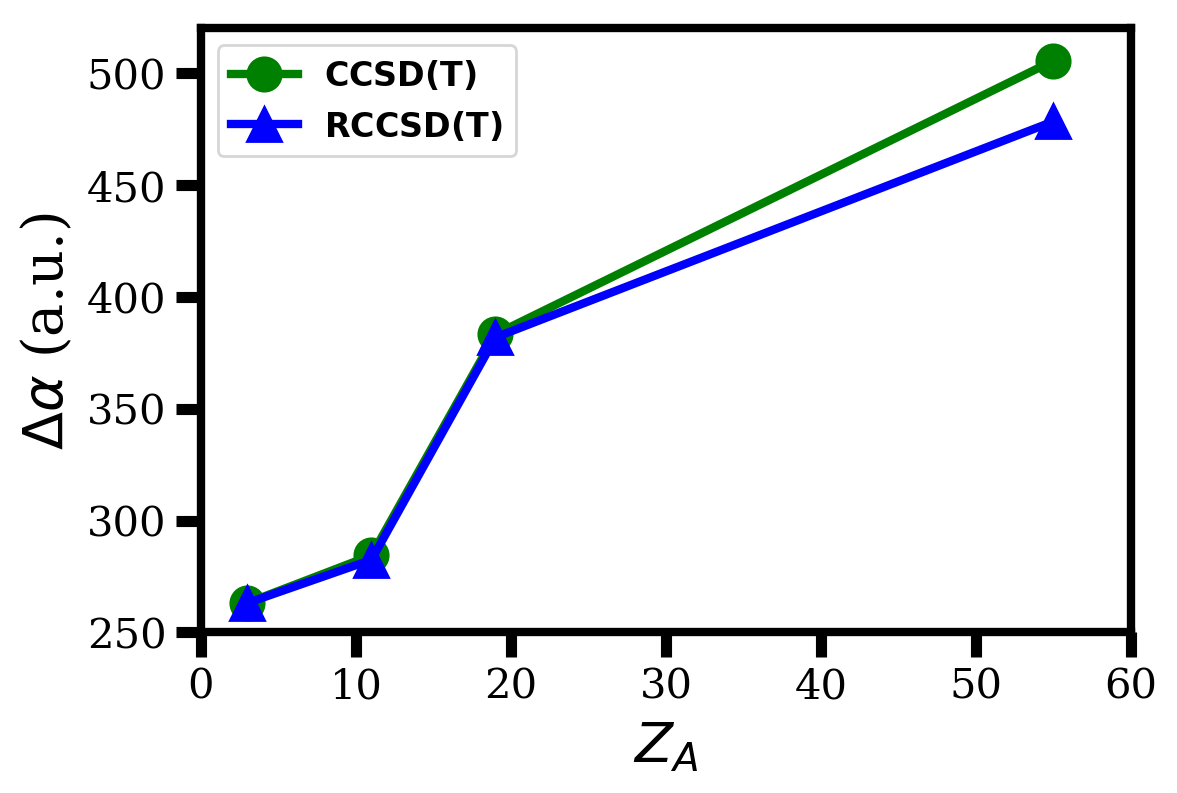} & \includegraphics[width=5cm, height=4.5cm]{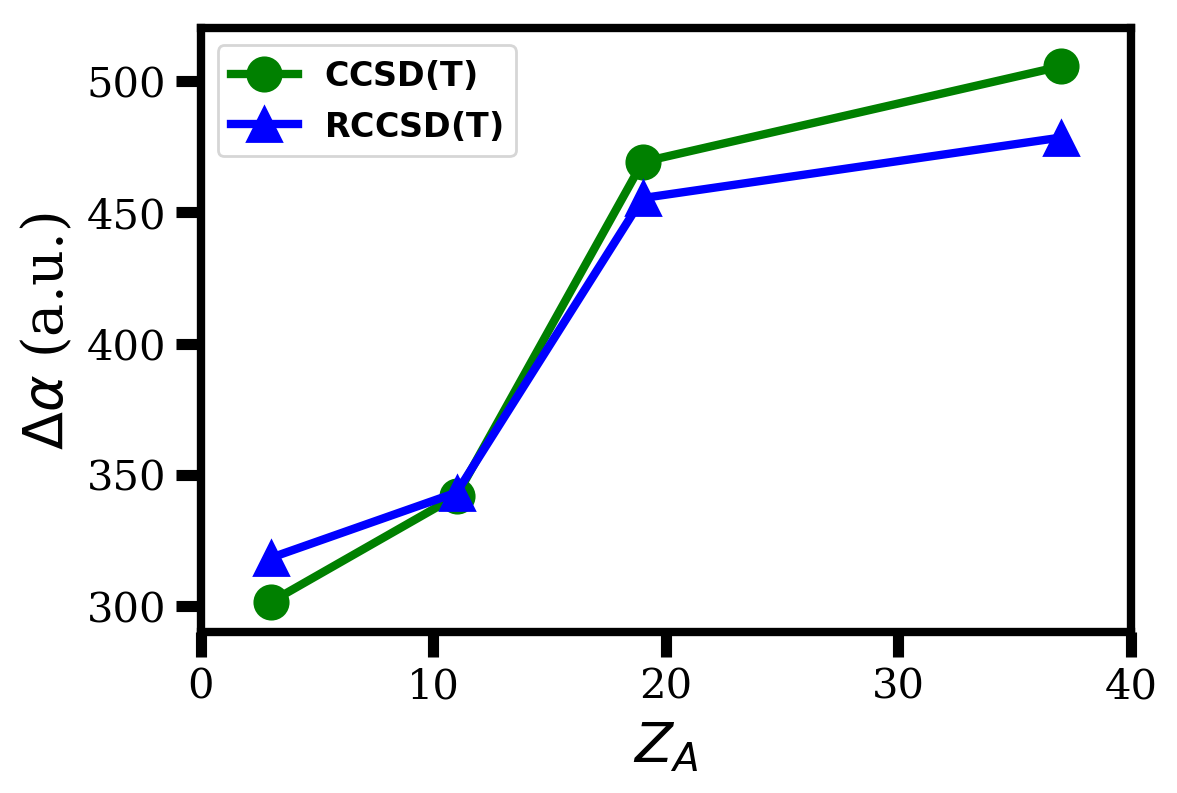} &\\
(d) & (e) & \\
\end{tabular}
\caption{(Colour online) Plots highlighting the variation of relativistic effects in the polarizability anisotropies (in a.u.) of the heteronuclear alkali dimers belonging to (a) Li-, (b) Na-, (c) K-, (d) Rb-, and (e) Cs-families using the CCSD(T) and RCCSD(T) methods.}
\label{fig:figure6}
\end{figure*}

\subsection{Results for polarizabilities}

We now turn to discussing parallel component results of polarizability. The $\alpha_{\parallel}$ values from our calculations, as well as those from previous works and experiments, are shown in Table~\ref{tab:table2}. It can be observed from this table that the effects of electron correlations are increasing the $\alpha_{\parallel}$ values from their HF and DF values, except in the cases of KRb, KCs, and RbCs. This is opposite to the trend that we observed for the PDM, where we found that inclusion of correlation effects lowered their magnitudes. We see that for KRb, KCs, and RbCs, the inclusion of partial triples increases $\alpha_{\parallel}$ in the non-relativistic calculations, while in the relativistic calculations, we observe a non-uniform trend where it increases $\alpha_{\parallel}$ values for KCs, and reduces it for KRb and RbCs. However, unlike in the case of PDM, the contribution from partial triples to $\alpha_{\parallel}$ is quite small, with NaCs differing the most between the (R)CCSD and (R)CCSD(T) results, at about 4 percent. Relativistic effects themselves do not become important for the molecules up to KRb, with $F$ being less than 2 percent throughout (at the CCSD(T) level of theory). However, the relativistic corrections result in a slightly higher $F$ of about 4.5 and 6.5 percent for KCs and RbCs, respectively. Figs.~\ref{fig:figure3}(a) through (e) reflect the smooth family-wise trend, where the relativistic effects steadily become important as the molecule becomes heavier, as expected. 

We now briefly discuss the previous works on the $\alpha_{\parallel}$ values of alkali dimers. There are no measurements of individual $\alpha_{\parallel}$ and $\alpha_{\perp}$ components; experiments obtain  average polarizability and polarizability anisotropy. We could only find limited works in literature that report calculation of $\alpha_{\parallel}$, and with the exception of Ref.~\cite{Zuchowski}, the other works discuss the polarizabilities of only one or a few alkali dimers. In Ref.~\cite{Urban}, Urban and Sadlej reported $\alpha_{\parallel}$ for LiNa, LiK, LiRb, NaK, NaRb, and KRb, using the CCSD(T) method (along with MVD corrections), with the electron correlations accounted from the valence and next-to-valence shells only. The authors in Ref.~\cite{2000}, on the other hand, employed a Numerov-Cooley (NC) scheme in their semi-numerical approach. They reported their results for $\alpha_{\parallel}$ of LiK in this approach by using Complete Active Space Self-Consistent Field (CASSCF) approach and  second-order Complete Active Space Perturbation Theory (CASPT2). They also perform CASSCF in combination with Bishop-Kirtman perturbation theory (BKPT), besides calculating vibrational corrections to $\alpha_{\parallel}$. Merawa \textit{et al}~\cite{2003} calculated $\alpha_{\parallel}$ of LiNa to be 350.6 a.u., by using the CCSD(T) method, and exciting all the electrons in their calculations. They also found this property to be 352.3 a.u., using a time-dependent gauge invariant (TDGI) method. The most recent work by Deiglmayr \textit{et al}~\cite{Deiglmayr} employed the CI approach by perturbing the  multi-configuration wave function, and had performed calculations on all the alkali dimers. We find that our RCCSD(T) results are in excellent agreement with their results up to NaCs (the differences are less than 2 percent), after which we observe a sharp deviation of up to 10 percent for KCs. We expect that the differences are not only because of relativity, but also due to correlation effects, recalling our observation that the electron correlations \emph{reduce} this quantity from the HF or DF to the (R)CCSD(T) methods only for these last three heavier molecules. 

\begin{figure}[t]
\includegraphics[width=8.5cm, height=6.0cm]{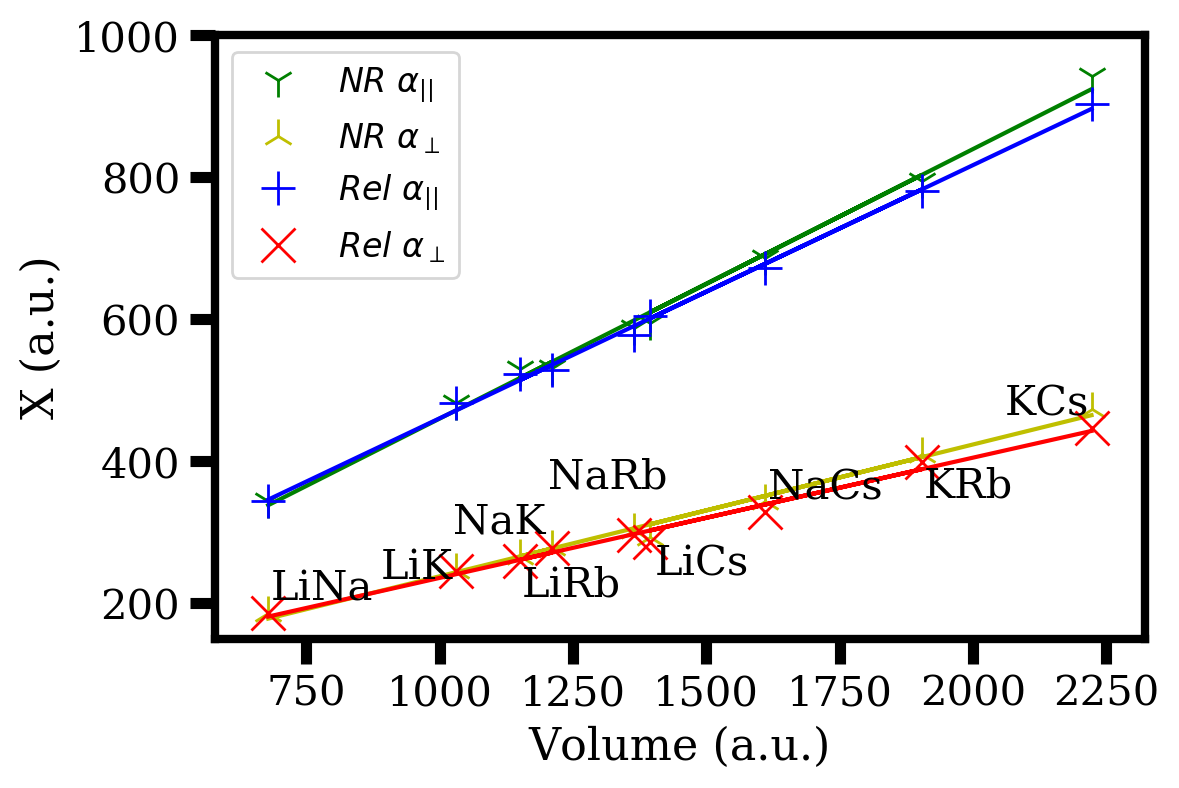}
\caption{(Colour online) The values of $\alpha_{\parallel}$ and $\alpha_{\perp}$ (commonly denoted by X, and given in a.u.) are plotted against volume (in a.u.). In the legend, `NR' refers to CCSD(T) results while `Rel' denotes to the RCCSD(T) values.}
\label{fig:figure7}
\end{figure}

Proceeding with the discussions on the results for $\alpha_{\perp}$, we find that these values consistently decrease with the inclusion of electron correlation effects in both the non-relativistic and relativistic calculations, as shown in Table~\ref{tab:table3}, contrary to $\alpha_{\parallel}$, and similar to PDM. Relativistic effects are also found to decrease $\alpha_{\perp}$. Examining $F$ values reflect that relativistic effects become more important as a molecule in a family gets heavier, with the exception of LiRb, as shown in Figs.~\ref{fig:figure4}(a) through (e). Also, similar to $\alpha_{\parallel}$, $F$ is the largest for RbCs, and is about 7 percent. The effect of partial triples is slightly higher for $\alpha_{\perp}$, and is about 4.5 percent for RbCs. In the previous paragraph, we had compared our calculated values for $\alpha_{\parallel}$ with those from earlier literature. All of those references also computed  $\alpha_{\perp}$ (with the exception of Ref.~\cite{2000}), and therefore, we will not discuss their methods again in this paragraph. 

We see from Table~\ref{tab:table4} that for the non-relativistic as well as the relativistic cases, the average polarizability value decreases from mean-field to (R)CCSD(T) methods, with the exception of LiNa. This can be understood from the fact that for LiNa, the parallel component increases more due to correlation than the decrease in the perpendicular component. This is not the case for all of the other molecules. In fact, the change in perpendicular components even dominates over that of parallel components for the heavier molecules. Relativity, as seen by comparing the CCSD(T) and RCCSD(T) results, further reduces $\bar{\alpha}$ results. Again, relativistic effects do not alter the average values of polarizabilities significantly, as they do not do so for the individual components. In Figs.~\ref{fig:figure5}(a) through (e), we show the trends in $\bar{\alpha}$ at the (R)CCSD(T) level that are arranged according to family. As seen, the relativistic effects within each family become increasingly important as the non-family atom (for example, in the Li family, it could be Na, K, Rb, or Cs) becomes heavier, with the exception of LiCs. The effect of partial triples is seen to be unimportant for $\bar{\alpha}$ from Table~\ref{tab:table4}. Our results also agree very well with the available experimental values, and are within the error bars of the measured values for NaK and KCs that are reported by Tarnovsky \textit{et al}~\cite{Tarnovsky}. These experimental results for NaK and KCs are obtained by combining measurements of average effective polarizabilities with the then-available PDM values taken from Igel-Mann \textit{et al}~\cite{Igel Mann}, at an average temperature of 612 K and 494 K for NaK and KCs, respectively. When we replaced the PDMs from Igel-Mann \textit{et al}~\cite{Igel Mann} with ours, we observed negligible difference in $\bar{\alpha}$ that Tarnovsky \textit{et al} obtained. However, to estimate $\bar{\alpha}$ for the other alkali dimers, the work by Tarnovsky \textit{et al} combines their measured homonuclear dimer polarizabilities with an empirical formula, thereby possibly introducing fairly large errors in some of their results. Regarding temperature dependency, we do not expect that our T$=0$ K results would differ significantly from the measurements carried out at the aforementioned temperatures, based on the earlier mentioned work by Muller and Meyer on homonuclear alkali dimers~\cite{Muller2}. In these rigorous studies, Muller and Meyer had shown that the dependency of average polarizability on a wide range of temperatures (between 0 and 1000 K) are not going to change the results significantly, and the maximum variation is anticipated to be about 10 percent from the values obtained at the zero temperature. 

For completeness, we also discuss the experimental results briefly for $\bar{\alpha}$ of LiNa from Ref.~\cite{Dag2}, where the authors have first measured polarizability anisotropy, $\Delta \alpha$, by determining Stark frequencies at some value of external electric field. They have combined this information with their knowledge of PDM (which is in turn obtained by measuring the rotational constant, as discussed in the previous paragraphs), to get $\bar{\alpha}$, as prescribed in Ref.~\cite{P. J. Dagdigian}. At this point, it is worthwhile to mention that all of the current theoretical values underestimate the average polarizability when compared with the experimental results for LiNa (we add that all of these theoretical values are within or very close to the error bars from experiment). This observation holds in spite of the theoretical results agreeing very well with their measurement for the anisotropy. However, since only one known experimental work exists (both for average polarizability and polarizbility anisotropy of LiNa), more detailed calculations and further experiments are possibly required, before arriving at further conclusions. 

The trends in $\Delta{\alpha}$ that are reflected in Figs.~\ref{fig:figure6}(a) through (e) stem from those in parallel and perpendicular components of polarizability, as $\Delta{\alpha}$ is the difference between $\alpha_{\parallel}$ and $\alpha_{\perp}$. For example, since relativistic effects increase $\alpha_{\parallel}$ while they decrease $\alpha_{\perp}$ for LiCs, we observe that relativity  matters the most for the molecule (about 5 percent). Also, partial triples in $\Delta{\alpha}$ become more important than in $\bar{\alpha}$, with NaCs and LiCs accounting for 8 and 9 percent, respectively. We are not discussing here the $\Delta{\alpha}$ results of Ref.~\cite{Dag2} as it has already been done in the previous paragraph. 

\begin{table}[t]
\caption{Demonstration of changes in $\alpha_\parallel$, $\alpha_\perp$, $\bar{\alpha}$, $\Delta\alpha$, and $\mu$ values of NaCs molecule at different virtual energy level cut-offs using the RCCSD(T) method. Calculations were performed using TZ basis functions. All the quantities are specified in a.u.. }
\centering
\begin{tabular}{lc ccccc}
\hline \hline
Cut-off &Method&	$\alpha_\parallel$&$\alpha_\perp$&$\bar{\alpha}$&$\Delta\alpha$&$\mu$ \\
\hline
 & & \\
1000&DF&621.78&385.80&464.46&235.98&2.42\\
1000&RCCSD&648.26&335.48&439.74&312.78&2.02\\
1000&RCCSD(T)&672.15&331.56&445.09&340.59&1.82\\ \\
2000&DF&621.75&385.81&464.46&235.94&2.42\\
2000&RCCSD&648.24&333.43&438.37&314.81&2.02\\
2000&RCCSD(T)&672.12&328.84&443.27&343.28&1.82\\ \\
5000&DF&621.78&385.81&464.47&235.97&2.42\\
5000&RCCSD&648.26&333.43&438.37&314.83&2.02\\
5000&RCCSD(T)&672.15&328.93&443.34&343.22&1.82\\
\hline \hline
\end{tabular}
\label{tab:table7}
\end{table}

\subsection{Volume effects}

Next, we address the dependence of the components of polarizability on volume. This aspect has been addressed by using models in the past, for example, see Refs.~\cite{Politzer,Korny}. It has also been discussed in Ref.~\cite{Deiglmayr}, where the volume, $V$, is defined as $\frac{4}{3}\pi r_e^3$, with $r_e$ denoting the equilibrium bond length. We plot the components of polarizabilities, both from non-relativistic and relativistic calculations, against volume in Fig.~\ref{fig:figure7}. From the figure, we see that a linear fit to our relativistic calculations gives $0.36V+104.26$ for $\alpha_{\parallel}$, and $0.17V+66.7$ for $\alpha_{\perp}$. We find that the ratio of the slopes of $\alpha_{\parallel}$ versus $V$ to $\alpha_{\perp}$ versus $V$ from our relativistic calculations agree well with the non-relativistic ones, and we obtain a value of about 2 for the ratio. This is in agreement with the slope obtained by Deiglmayr \textit{et al}~\cite{Deiglmayr} from their calculations. The linear polarizability-volume relationship could be viewed as an effective elliptic charge distribution for a dimer at a specified $r_e$~\cite{Deiglmayr}. Although relativistic and non-relativistic results for the lighter systems are very close to each other, we observe that the linear fits between the two cases deviate further as we go to the heavier molecules, while continuing to preserve the ratio of the slopes. 

\subsection{Error Analysis}

We now discuss the possible sources of uncertainties in our calculations of PDMs and polarizabilities of the considered alkali dimers. We assume our RCCSD(T) values are the most accurate among other methods in literature and therefore the uncertainties are estimated for these results. Since we have adopted the FF approach, it is essential to choose the perturbation parameter, $\mathcal{E}$, carefully in order to obtain reliable results. Our choice of $\mathcal{E}=10^{-4}$ a.u. is consistent with those from the previous works that had determined PDMs and  polarizabilities using the FF procedure. However, we had also verified consistencies in the results by performing calculations of PDMs as well as polarizabilities by using other values of $\mathcal{E}$, namely $10^{-3}$, $5\times 10^{-4}$, and $10^{-4}$ a.u.. For this purpose, and in view of minimizing the computational cost, we chose only the Li family as a representative case, and repeated the non-relativistic calculations with a double-zeta (DZ) basis. We did not find any significant differences in these calculations due to the choice of $\mathcal{E}$. We also anticipate similar trends with the relativistic calculations and in other heavier alkali dimers. We found that the PDM values hardly change, while the parallel and perpendicular components of polarizability smoothly change in the first decimal place for LiNa and LiK, and within 3 a.u. for LiRb and LiCs. Also, the truncation errors that could result from numerical differentiation schemes have been taken care of by comparing our results using three-point as well as five-point formulae and we found that the results in both those approaches are identical. 

It is also imperative to ensure that there is negligible uncertainty involved due to cut-off of virtual orbitals in our relativistic RCCSD(T) calculations for the heavier alkali dimers. Therefore, we chose NaCs, a moderately heavy molecule where relativistic effects are sufficiently important and yet practical for multiple computations, for this purpose. The results with different set of virtual orbitals are tabulated in Table~\ref{tab:table7}. It shows that the PDM values remain identical, while the components of polarizability change in the second decimal place, which are much smaller than the level of accuracy intended to achieve in the present work. 

We now move on to discussions on the most traditional uncertainties due to neglected effects in our calculations. It is beyond the scope of this work to estimate contributions due to the full Breit and quantum electrodynamics interactions. However, it is expected that these higher-order relativistic corrections will not exceed 0.5\% in all the considered molecules. Uncertainties could also arise because of neglecting contributions from higher level excitations in the (R)CC theory and use of incomplete basis functions. The percentage fraction difference in our relativistic results from RCCSD to RCCSD(T) methods indicate that higher-level excitations should not contribute beyond 5 percent to the PDMs. A similar analysis provides us with an error estimate of about 3 percent for $\alpha_{\parallel}$ and $\alpha_{\perp}$. We now analyse the error due to incompleteness in basis. We had employed a TZ basis for our relativistic calculations, and included diffuse functions, wherever available. We could not, however, perform relativistic calculations using a QZ basis, as they are forbiddingly expensive, even for moderately heavy systems like KRb. Therefore, we resort to an approximation, where we first perform extensive CBS calculations for the alkali dimers using the CCSD(T) method. We employed the two-point scheme by Helgaker~\cite{cbs0,cbs1} for CBS, which was mentioned earlier. We then approximate the relativistic CBS value of $\mu$ and $\alpha$ (commonly denoted here as $O$ for ease in notation) by $O_{CBS}^{Rel} \approx (O_{CBS}^{NR}/O_{TZ}^{NR})O_{TZ}^{Rel}$, where the subscripts refer to the basis and the superscripts indicate whether the property has been obtained using an non-relativistic calculation or a relativistic one. With this approximation, we obtain a percentage fraction difference of less than 2 percent for the PDMs of the alkali dimers, with the exceptions of LiNa, NaK, and NaRb, as the PDMs for these systems do not converge from DZ through quadruple-zeta (QZ) basis, hence making a CBS extrapolation not possible. However, we do not expect the errors to be beyond 2 percent in these cases too. A similar exercise was also carried out for $\alpha_{\parallel}$, and we found that the fraction difference was less than 4 percent for all the alkali dimers, except for LiRb. We also verified our approximate formula for $O_{CBS}^{Rel}$, by explicitly performing RCCSD(T) calculations for the PDM and $\alpha_{\parallel}$ of NaCs (with aug-cc-pCVTZ for Na and Dyall's 4Z basis for Cs), and obtained exactly the same PDM as that from $O_{CBS}^{Rel}$, while $\alpha_{\parallel}$ differed from the $O_{CBS}^{Rel}$ estimate by just 0.6 percent. We would expect an error percentage that is similar to that in $\alpha_{\parallel}$ for the perpendicular components of polarizabilities too, which is at most 4 percent. Finally, we linearly add the errors and estimate that the uncertainties in our relativistic calculations are about 8 percent for the PDMs, as well as for the parallel and perpendicular components of polarizabilities. 

\begin{table}[t]
\centering
    \caption{Improved values of isotropic $C_6$ coefficients (in a.u.) by combining our estimated $C_6^{ind}$ and $C_r^{rot}$ contributions with the $C6^{disp}$ contributions borrowed from Ref.~\cite{Zuchowski}. We have also compared these results with previously reported two non-relativistic calculations. The differences between our results with other calculations demonstrate importance of relativistic calculations in the determination of $C_6$ coefficients.}
    \label{tab:table6}
\begin{tabular}{l@{\hspace{0.6in}}c@{\hspace{0.6in}}l}
\hline \hline
Molecule&Reference&$C_6$ value\\
\hline
 & & \\
LiNa&Ref.~\cite{Bohn}&3\ 880\\
&Ref.~\cite{Lepers}&3\ 583\\
&Ref.~\cite{Zuchowski}&3\ 709\\
&This work &3 807\\
LiK&Ref.~\cite{Bohn}&524\ 000\\
&Ref.~\cite{Lepers}&570\ 190\\
&Ref.~\cite{Zuchowski}&411\ 682\\
&This work &434 316\\
LiRb&Ref.~\cite{Bohn}&1\ 070\ 000\\
&Ref.~\cite{Lepers}&1\ 252\ 300\\
&Ref.~\cite{Zuchowski}&884\ 705\\
&This work &929\ 144\\
LiCs&Ref.~\cite{Bohn}&3\ 840\ 000\\
&Ref.~\cite{Lepers}&4\ 585\ 400\\
&Ref.~\cite{Zuchowski}&3\ 409\ 406\\
&This work &3 664 836\\
NaK&Ref.~\cite{Lepers}&561\ 070\\
&Ref.~\cite{Zuchowski}&516\ 606\\
&This work &518 370\\
NaRb&Ref.~\cite{Lepers}&1\ 524\ 900\\
&Ref.~\cite{Zuchowski}&1\ 507\ 089\\
&This work &1\ 457\ 076\\
NaCs&Ref.~\cite{Lepers}&7\ 323\ 100\\
&Ref.~\cite{Zuchowski}&6\ 946\ 696\\
&This work &7\ 086\ 877\\
KRb&Ref.~\cite{Lepers}&15\ 972\\
&Ref.~\cite{Zuchowski}&17\ 720\\
&This work &17\ 542\\
KCs&Ref.~\cite{Lepers}&345\ 740\\
&Ref.~\cite{Zuchowski}&469\ 120\\
&This work &469\ 769\\
RbCs&Ref.~\cite{Lepers}&147\ 260\\
&Ref.~\cite{Zuchowski}&180\ 982\\
&This work &190\ 442\\
\hline \hline
    \end{tabular}
\end{table}

\subsection{Implications on determining $C_6$ coefficients}

We intend to discuss here an important application of our results apart from their general demand to carry out relativistic calculations. As known, when two heteronuclear alkali dimer molecules interact via a long-range Van der Waals interaction, its dominant potential is given by $-C_6/r^6$ \cite{review,Zuchowski,Bohn}. Here, $r$ is the inter-molecular separation and $C_6$ is known as Van der Waals coefficient. For molecules, $C_6$ can be expressed as \cite{review,Zuchowski,Bohn}
\begin{eqnarray}
C_6 = C_6^{disp}+C_6^{ind}+C_6^{rot},
\end{eqnarray}
where the three terms on the right hand side are known as the dispersion (denoted by superscript, `disp'), the induction (denoted by superscript, `ind'), and the rotational (denoted by superscript, `rot') contributions, respectively. We estimate the induced contribution, using the expression \cite{Zuchowski}
\begin{eqnarray} 
C_6^{ind} = 2 \mu^2\bar{\alpha} , 
\end{eqnarray}
by substituting our calculated PDM and $\bar{\alpha}$ values. 
Similarly, we determine the rotational contributions using the expression given by
\cite{review,Zuchowski,Bohn}
\begin{eqnarray}
C_6^{rot} = \frac{\mu^4}{6B} .
\end{eqnarray}
Due to the fourth-power dependence on PDM, the rotational term dominates over the sum of the other two terms by at least an order of magnitude in the evaluation of $C_6$ values for molecules with large PDMs \cite{review}. This is indeed the case for eight of the ten alkali dimers with the exceptions LiNa and KRb, owing to their small PDMs and larger $B$ value of LiNa. This dependence on accurate calculations of $C_6$ coefficients become more relevant for molecules such as LiCs and KCs, for which experimental values of PDM do not exist. For estimating $B$ values, we consider the $^7$Li, $^{23}$Na, $^{41}$K, $^{87}$Rb, and $^{133}$Cs bosonic isotopes. 

We, however, have borrowed the most accurately calculated results for the dispersion terms from Ref.~\cite{Zuchowski}. This is done keeping in mind that the dispersion contributions are at least one order lesser than the rotational ones for most of the alkali dimers. We tabulate all these contributions and the final results of $C_6$ for various heteronuclear dimers in Table~\ref{tab:table6}. It can be clearly seen from this table that use of revised $C_6^{ind}$ and $C_6^{rot}$ contributions change the final results of $C_6$ significantly than the values reported in Ref. \cite{Zuchowski}. In fact, the results become substantially different compared to pure non-relativistic calculations of Ref. \cite{Bohn}, which are also quoted in the above table for the comparison. We see from the table that the isotropic $C_6$ coefficient can vary as much as 7 percent for LiCs, when compared to that from Ref.~\cite{Zuchowski}, while it can be about 20 percent for LiK and 15 percent for LiRb with respect to Ref.~\cite{Bohn}, when relativistic effects are included in obtaining the PDM and polarizabilities. We also observe that there are significant differences between our results and those obtained from the recent calculations in Ref.~\cite{Lepers}, and are over 25 percent for LiK, LiRb, LiCs, and KCs. At this point, we would also like to draw attention to the fact that although PDM values from Ref.~\cite{Zuchowski} are in close agreement with ours, the differences in our results are still sufficiently large to lead to a non-negligible change in $C_6$ values owing to the $\mu^4$ dependence. This clearly highlights the crucial roles that accurate calculation of the PDM plays in determining the $C_6$ coefficients of alkali dimers. 

\section{Conclusion} \label{sec4}

In summary, we have performed four-component relativistic finite-field calculations of both the permanent electric dipole moments as well as static dipole polarizabilities of heteronuclear alkali dimers in their ground states using the coupled-cluster theory and compared these results with the non-relativistic calculations at the same level of approximations. We observe that the relativistic effects become very important for the determination of permanent electric dipole moment values, especially in the heavier molecules. To this end, we address the long-standing issue in the permanent electric dipole moment of LiNa, where calculations are at disagreement with measurements. We do so by presenting improved calculation of the permanent electric dipole moment of LiNa using complete basis set extrapolation, which agrees very well with the most precise experimental value. We also discuss the importance of analyzing the trends in electron correlation effects based on different groups of molecules that we categorise in terms of family. We  compare our results with the previous experimental and theoretical works. We discuss the variation of the components of dipole polarizability with volume. We present possible sources of uncertainties in our calculations of the above quantities. Further, we demonstrate the importance of considering relativistic effects in the determination of the permanent electric dipole moments and static dipole polarizabilities, by using them in evaluating the Van der Waals $C_6$ coefficients for the alkali dimers. 

\section*{Acknowledgments}

All the computations were performed on the VIKRAM-100 cluster at PRL, Ahmedabad. We thank Prof. Trond Saue for providing useful insights on the symmetry aspects of the Dirac program. We also thank Mr. Madhusudhan. P for many discussions on molecular alignment.

\end{document}